# Direct and Simultaneous Observation of Ultrafast Electron and Hole Dynamics in Germanium

Michael Zürch[1,†,*], Hung-Tzu Chang[1,†], Lauren J. Borja[1,†], Peter M. Kraus[1], Scott K. Cushing[1], Andrey Gandman[1,‡], Christopher J. Kaplan[1], Myoung Hwan Oh[1,2], James S. Prell[1,§], David Prendergast[3], Chaitanya D. Pemmaraju[3,4], Daniel M. Neumark[1,5,*], Stephen R. Leone[1,5,6,*]

[1] Department of Chemistry, University of California, Berkeley, CA 94720, USA

[2] Materials Sciences Division, Lawrence Berkeley National Laboratory, Berkeley, CA 94720, USA

[3] The Molecular Foundry, Lawrence Berkeley National Laboratory, Berkeley, CA 94720, USA

[4] Stanford Institute for Materials & Energy Sciences, SLAC National Accelerator Laboratory, Menlo Park, CA 94025, USA

[5] Chemical Sciences Division, Lawrence Berkeley National Laboratory, Berkeley, CA 94720, USA

[6] Department of Physics, University of California, Berkeley, CA 94720, USA

[‡] Present address: Solid State Institute, Technion – Israel Institute of Technology, Haifa, 32000, Israel

[§] Department of Chemistry and Biochemistry, University of Oregon, Eugene, OR 97403, USA

[*] Correspondence should be addressed to: mwz@berkeley.edu (M. Z.); dneumark@berkeley.edu (D. M. N.); srl@berkeley.edu (S. R. L.)

[†] These authors contributed equally to this work.

**Understanding excited carrier dynamics in semiconductors is crucial for the development of photovoltaics and efficient photonic devices. However, overlapping spectral features in optical/NIR pump-probe spectroscopy often render assignments of separate electron and hole carrier dynamics ambiguous. Here, ultrafast electron and hole dynamics in germanium nanocrystalline thin films are directly and simultaneously observed by attosecond transient absorption spectroscopy (ATAS) in the extreme ultraviolet at the germanium $M_{4,5}$-edge ($\sim 30$ eV). We decompose the ATAS spectra into contributions of electronic state blocking and photo-induced band shifts at a carrier density of $8 \times 10^{20}$ cm$^{-3}$. Separate electron and hole relaxation times are observed as a function of hot carrier energies. A first order electron and hole decay of $\sim 1$ ps suggests a Shockley-Read-Hall recombination mechanism. The simultaneous observation of electrons and holes with ATAS paves the way for investigating few to sub-femtosecond dynamics of both holes and electrons in complex semiconductor materials and across junctions.**

Investigation of the ultrafast photoexcited electronic response in semiconductors has provided invaluable insights into the carrier dynamics and dielectric properties of many materials. With the advent of femtosecond laser techniques, electron and hole scattering in semiconductors has been observed individually and characterized by optical and infrared (IR) pump-probe experiments[1,2]. However, the time-resolved observation and characterization of electron and hole kinetics simultaneously in the optical regime is challenging due to overlapping spectral signatures requiring narrow-band excitation to separate pump from probe, which inherently limits the temporal resolution. We address this issue using attosecond transient absorption in the extreme ultraviolet region of the spectrum, with specific application of the method to electron and hole dynamics in photoexcited germanium.

Attosecond transient absorption spectroscopy (ATAS) in the extreme ultraviolet (XUV) is an important technique for studying electron dynamics at the sub-fs and few fs timescale[3,4]. ATAS has been successfully applied to investigate the dielectric response of insulators[5,6] and carrier dynamics in semiconductors[7,8] down to sub-fs time scales. In ATAS, a visible-to-near infrared (VIS-NIR) pump pulse excites carriers first, and after a given time delay τ, a broadband attosecond pulse in the XUV generated by high harmonic generation (HHG)[9] excites core-level electrons into the valence and conduction bands. The spectrally resolved transient absorption of the XUV encodes the dynamics of excited carriers and possible band modifications in the semiconductor. The core-level excitation with XUV photons is element specific, making the technique advantageous for investigating carrier dynamics in heteroatomic, ternary, and quaternary systems. The photon energy of the broadband XUV pulse lies far above the interband excitation energies, and therefore ATAS in principle allows simultaneous capture of electron and hole dynamics without further inducing interband carrier excitations.

Germanium is a Group IV semiconductor with an indirect band gap of 0.66 eV. Previous optical experiments observed electron scattering from one part of the band structure to another, i.e. intervalley scattering, occurring on a few hundreds of fs time scale[10–13]. In separate experiments, interband scattering of holes has been investigated[14,15]. Recently, studies of the intervalley scattering of carriers in germanium have attracted renewed interest due to the prospect of spin-polarization induced in the scattering event, which is useful for the development of spintronics[15–18]. In this paper, measuring XUV transient absorption at the germanium $M_{4,5}$-edge spectrally resolves signatures of electrons and holes simultaneously, which enables the time- and energy-resolved tracking of electron and hole kinetics. In the present experiment, carrier-induced and heat-induced band shifts are observed in addition to state-filling. With ATAS, the electron and hole relaxation and recombination times within a single experiment on nanocrystalline germanium thin films are successfully characterized.

This contribution is structured as follows. In the following section the experiment is introduced. Subsequently, we establish a model that allows decomposing the measured raw transient absorption data into contributions of state blocking, shifts of the excited state spectrum and broadening. We show how the carrier dynamics can be extracted by separating spin-orbit components in the XUV. Using the retrieved carrier dynamics, we assign the valence and conduction band states and establish that electrons and holes are simultaneously observed by comparing to first principles theory calculations. This is followed by analysing the kinetics of the electrons in the conduction band (CB), the hole kinetics in the valence band (VB), and the band shift dynamics, which are dominant. In the discussion the observed kinetics of the two charge carriers are addressed to synthesize a complete picture of the dynamics.

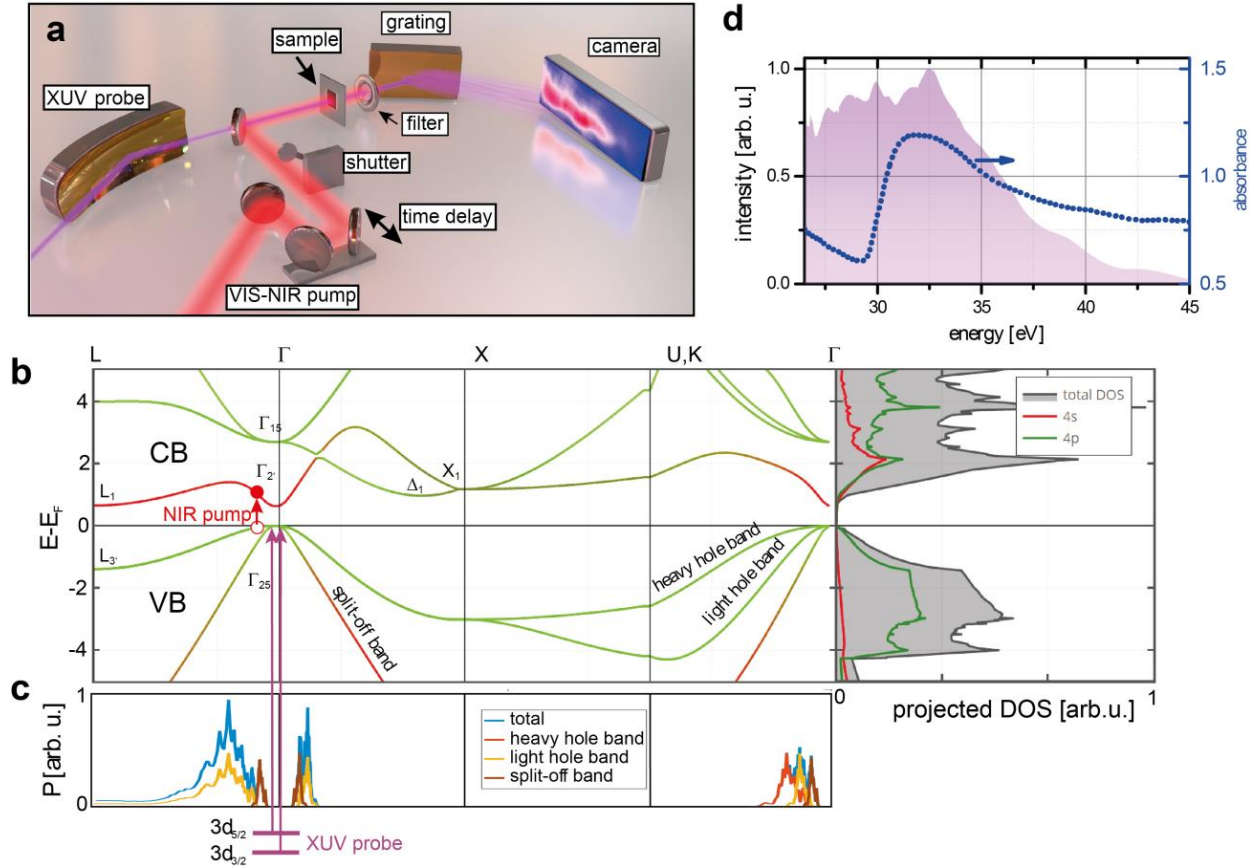

**Figure 1. XUV attosecond transient absorption spectroscopy in germanium.** (a) In the experiment a time-delayed broadband XUV pulse is used to probe the transient absorption of a nanocrystalline germanium thin film after excitation with a broadband visible-to-near infrared (VIS-NIR) pump pulse. (b) Band diagram and projected density of states for germanium, calculated here by density functional theory calculation (see Methods). The VIS-NIR pump pulse initially promotes electrons (filled red circle) into the CB leaving behind a hole (open red circle) in the VB. (c) Transition probability $P$ for the specific VIS-NIR pump pulse used in this experiment at different parts of the band diagram (see Supplementary Information). The large bandwidth of the pump pulse allows to generate holes at all slopes of the Γ valley in the light hole band (yellow solid

line in (c)) and heavy hole band (red solid line in (c)) as well as the split-off band (brown solid line in (c)) by one-photon transitions without the assistance of phonons (i.e. direct transitions). At the $M_{4,5}$-edge the XUV pulse probes the transient state in the VB and CB from a *3d* core-level as indicated by the purple lines and arrows in panels (b) and (c); the *3d* core level has a significant spin-orbit splitting in germanium of 0.58 eV (ref. 20), and transitions from both spin orbit states are observed to those parts of the bands that are of *4p* orbital character. The valence band (VB) and conduction band (CB) in germanium are primarily of *4s* and *4p* orbital character. The orbital character is encoded by a red and green colour code, respectively, for *4s* and *4p* orbital character in panel (b). (d) The spectrum of the broadband attosecond probe pulse covers the $M_{4,5}$-edge of germanium (see absorbance in dotted blue line in (d)) around 30 eV.

**Attosecond transient absorption experiment on nanocrystalline germanium**

In Fig. 1 the XUV transient absorption experiment performed on a nanocrystalline germanium film is illustrated schematically. In the experimental apparatus (Fig. 1a) a time-delayed VIS-NIR pump pulse is collinearly superimposed with a broadband XUV pulse and focused onto the sample. The transient absorption signal $\Delta A_{\text{meas}}(E, \tau) = A_p(E, \tau) - A(E)$ is defined by the difference between the absorbance of the excited (pumped) thin film $A_p$ and the static absorbance $A$. Depending on the time delay $\tau$ between the VIS-NIR excitation (pump) pulse and XUV probe pulse the transient absorption as a function of wavelength and time is recorded by spectrally resolving the XUV light after the sample using a spectrometer. A detailed explanation of the experimental apparatus and sample characterization is in the Methods Section and Sec. S1 in the Supplementary Information. The samples are produced by electron beam deposition and annealing, leading to nanocrystalline domains, most likely containing significant defects. The few-cycle VIS-NIR pulse spanning 1.2-2.5 eV photon energy promotes electrons across the band gap from the VB into the CB (Fig. 1b). Given the spectrum of the exciting VIS-NIR pulse, the region of possible carrier excitation within the Brillouin zone of germanium is calculated (see Sec. S2 in Supplementary Information). Fig. 1c summarizes the possible optical excitations from the heavy-hole, light-hole, and split-off bands to the four low-lying conduction bands. All significant photon energies contained in the VIS-NIR pulse allow for a direct one-photon transition between the bands, hence indirect transitions assisted by phonons are expected to play a minor role for the pump pulse used[19]. The excitation pulse mainly excites carriers within the Γ valley and the XUV transient absorption tracks the intervalley and intravalley scattering, i.e. the thermalization of holes and electrons, and ultimately the recombination of the carriers. The XUV attosecond pulse, whose spectrum spans the germanium $M_{4,5}$ absorption edge (Fig. 1d), allows to probe an electron originating from a germanium $3d$ core-level to a respective VB or CB state. In general, the XUV pulse can probe the transient populations in both the CB and VB as both are of partial *4p* orbital character (green colour code in Fig. 1b), while transitions to parts of the bands that are of *4s* character are forbidden by dipole selection rules (red colour code in Fig. 1b) and invisible to the probe. The instrumental response time has an upper bound estimated to be ~7 fs (see Methods).

## Results

**Decomposition of transient absorption spectra into state blocking, broadening and spectral shifts**

Raw transient absorption spectra at different time delays $\tau$ are shown in Fig. 2a. The sample was illuminated using a VIS-NIR pulse with an intensity of $2.0 \times 10^{11}$ W/cm$^2$ exciting approximately $8 \times 10^{20}$ carriers per cubic centimetre or ~0.3% of the total number density (see Methods), which constitutes a high carrier density regime where carrier-carrier scattering effects are expected to play a role. In the time delay span measured from -50 fs to 1.5 ps, where negative delay means the XUV pulse arrives first, transient signals are observed for all time delays. Several features with positive and negative signs in $\Delta A$ are observed in the transient absorption signal with most of the observable dynamics taking place during the first picosecond following excitation. Overlapping spectral features arise from the spin-orbit splitting of the $3d_{5/2}$ and $3d_{3/2}$ core-level transitions, which has an energy splitting of 0.58 eV, comparable to the band gap (0.66 eV)[20]. It is expected to observe dynamical features of band shifts due to the excited electron-hole plasma[21] and phonon heating of the lattice[22,23], carrier excitation induced core-level shifts and a spectral broadening of the $M_{4,5}$-edge (Fig. 2b)[24], and electronic state blocking from excited carriers. As a whole, the complexity of underlying effects that contribute to the observed transient absorption spectra require disentangling these from the raw data to study the contributions individually.

We introduce an iterative procedure that takes possible spectral broadening, shifting and state blocking into account and seeks to break up these individual contributions from the measured raw data. In the iterative procedure, the individual contributions by shifting and broadening can be calculated from the measured static absorbance $A(E)$ (Fig. 2b) of the sample using a linear energy shift $\Delta E_{\text{shift}}$ and by convoluting the absorbance with a Gaussian of width $\sigma$, respectively. The state blocking can be estimated in the first iteration by two Gaussians having a width corresponding to the bandwidth of the VIS-NIR pulse with opposite signs for electrons and holes spaced by the central energy of the VIS-NIR pulse and subsequently by considering the spin-orbit splitting of the core-level. The assumption of a Gaussian shape of the state blocking is chosen as reasonable compromise between the expected initial distribution related to the NIR-VIS spectrum, which transitions into a Fermi-Dirac type distribution after rapid thermalization, convoluted with the instrumental resolution of the instrument. The time dependence of the state blocking is first estimated as a single exponential decay with a characteristic time constant retrieved from the experimental data. By minimizing an error metric taking the sum of these components and the measured signal into account, in the first iteration the underlying time dependent shift $\Delta E_{\text{shift}}(\tau)$ and broadening $\sigma(\tau)$ are estimated. As a convention, here a positive value of $\Delta E_{\text{shift}}(\tau)$ means a redshift of the excited state spectrum. In subsequent iterations the state blocking (SB) component $\Delta A_{\text{SB}}(E,\tau)$ is refined along with refinement of $\Delta E_{\text{shift}}(\tau)$ and $\sigma(\tau)$, converging after five iterations. It is found that broadening plays only a minor role in the experiments presented here and constitutes signals at the noise level (Fig. 2d). The major contributions arise from a time-dependent redshift (Fig. 2e) and the state blocking (Fig. 2c) following carrier excitation. A detailed description of the iterative procedure is given in Sec. S3 in the Supplementary Information.

As noted, the state blocking transient obtained (Fig. 2c) contains contributions from both $3d_{5/2}$ and $3d_{3/2}$ core-level transitions. A Fourier reconstruction method (see Methods and Sec. S4, Supplementary Information) is applied to the transient absorption signal to retrieve the contribution from a single spin-orbit state transition ($3d_{5/2}$). Figure 2f shows the spin-orbit-separated state blocking transient $\Delta A_{SB,3d_{5/2}}(E,\tau)$ that we will subsequently refer to as *carrier dynamics*.

In the time-dependent band shift $\Delta E_{shift}(\tau)$ (Fig. 2g) at negative time delays a constant redshift is measured, which originates from a heat-induced band shift owing to the thin film being heated up in the multipulse exposure experiment. This transient feature allows retrieving the temperature of the thin film, which ranges from 325 to 475 K for the data presented here (see Methods and Sec. 5 in Supplementary Information).

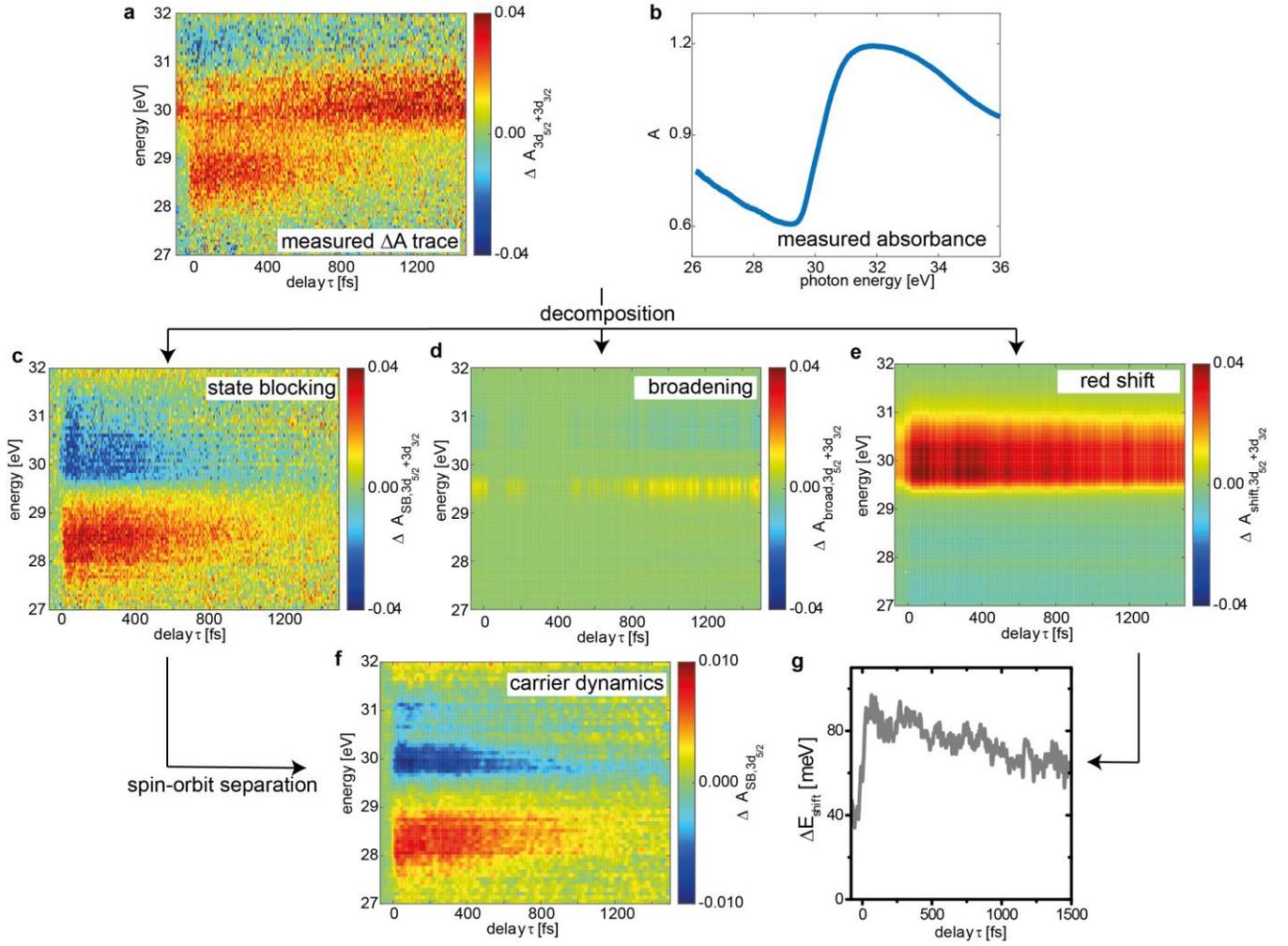

**Figure 2. Decomposition of the contributions from state blocking, broadening, and band shifts.** (a) Raw transient absorption data. Using the measured static absorbance (b) the measured $\Delta A_{\text{meas}}$ trace (a) is decomposed into 3 major components: (c) state blocking, (d) broadening of the excited state and (e) a redshift of the ground state, via an iterative algorithm, see text for details. (f) Modifying the state blocking contribution with a subsequent spin-orbit separation allows quantitative visualization of the electron and hole contributions, referred to as carrier dynamics. (g) The amount of redshift $\Delta E_{\text{shift}}(\tau)$ over the delay shows a constant non-zero shift for negative time delays, which is heat induced from previous laser pulses, and a time dependence for positive delays.

**Feature assignment in the experimental transient absorption data**

Fig. 3a shows the retrieved carrier dynamics $\Delta A_{SB,3d_{5/2}}(E,\tau)$ (same as Fig. 2f) with assignments of the valleys in germanium (cf. Fig. 1b). In general, two features are apparent, a broad positive feature around 28.2 eV and a broad negative feature spanning 29.5 to 31.0 eV. The signs of the carrier dynamics features can be understood as follows. With VIS-NIR excitation of electrons from VB to CB, holes are created in the VB, resulting in an increase of XUV absorption below the $M_{4,5}$-edge. Oppositely, the VIS-NIR-excited electrons in the CB block XUV transitions from the core-level, leading to a decrease in XUV absorption and, thus, a negative $\Delta A$. The carrier dynamics features as well as the band shift $\Delta E_{shift}(\tau)$ rise within the instrumental response time (Fig. 3b) closely following the exciting VIS-NIR pulse. The slow rise of the signal before time zero can be associated with a small pedestal or pre-pulse, for which the allowed one-photon transition in this experiment is susceptible to observation, in contrast to previous semiconductor attosecond experiments that used nonlinear excitation schemes[7,8].

To support the assignment of the features in the carrier dynamics signal, the profiles averaged over time delays between +8 to +12 fs (black rectangle in Fig. 3a) are compared to differential absorption profiles obtained by first-principles calculations (Fig. 3c). Here, real-time time-dependent density functional theory (TDDFT) was first employed to calculate the electronic excitation of crystalline germanium by a VIS-NIR pulse comparable to the one used in the experiments (see Methods). From the retrieved populations in the CB and VB following excitation, an excited-state X-ray absorption spectrum (XAS) was then calculated to obtain a transient absorption spectrum (dashed lines in Fig. 3c) with respect to a calculated XAS ground state spectrum, where both spectra are calculated for a single spin-orbit state. The comparison of the calculated (dashes in Fig. 3c) and experimental transient absorption (blue solid area in Fig. 3c) strongly corroborates the assignment of electrons and holes to the negative and positive carrier dynamics feature, respectively. This assignment is further supported by the binding energy of the $3d_{5/2}$ core-level at 29.2 eV coinciding with the gap between the two main features relating to the band gap.

Given that TDDFT assumes a crystalline unit cell, compared to the possible defect-prone nanocrystalline sample, the experimental and theory curves show excellent agreement.

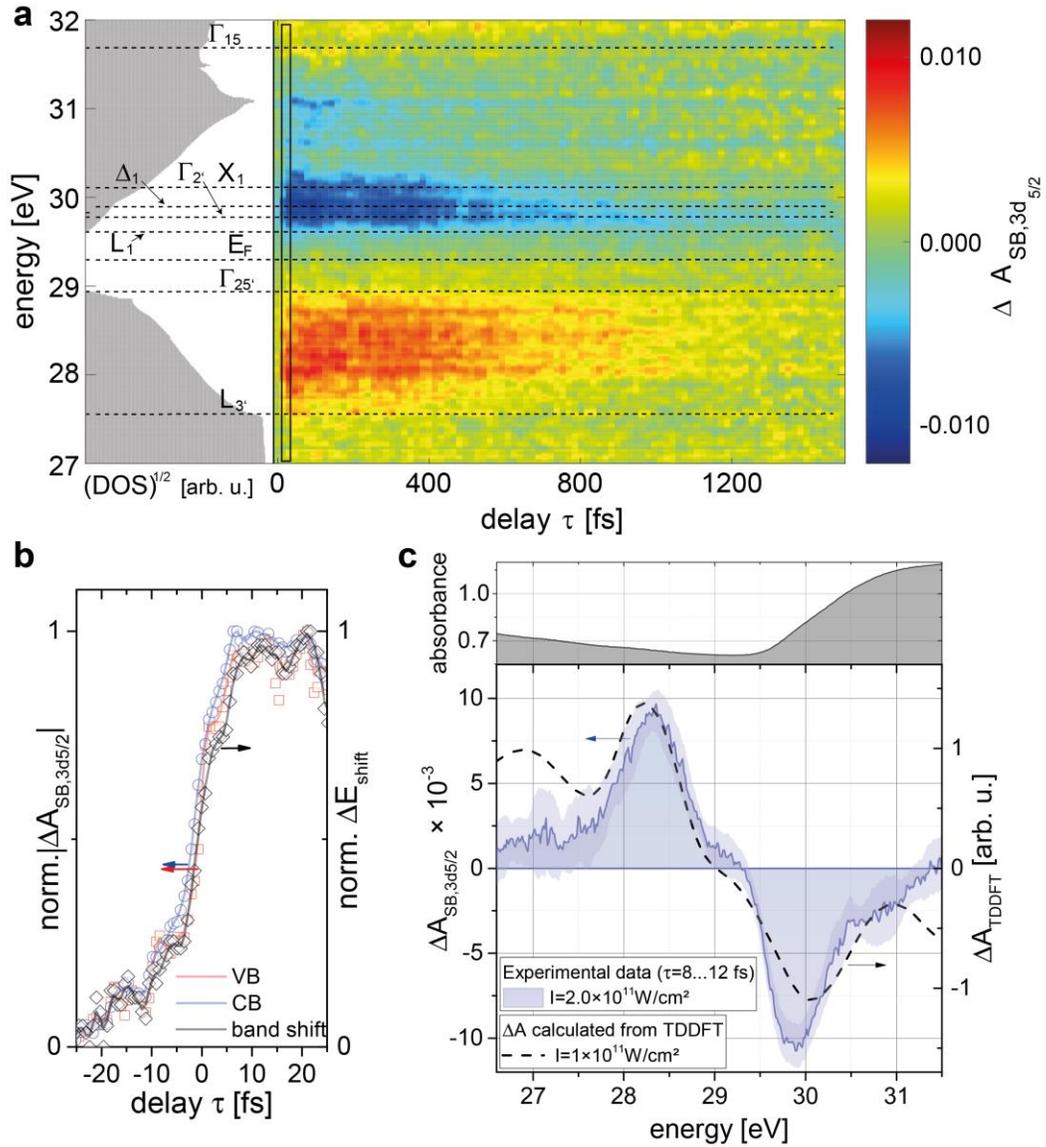

**Figure 3. Differential absorption experiment in comparison to first principles calculations.** (a) A carrier dynamics signal $\Delta A_{SB,3d_{5/2}}(E,\tau)$ (I=2×10$^{11}$ W/cm²) features positive and negative differential absorption in the VB and CB, respectively. Positive time delays correspond to the VIS-NIR pump pulse arriving before the XUV probe pulse. Comparison to a calculated density of states (DOS) allows assigning characteristic valleys of the band structure to the measured energy axis (cf. Fig. 1c). (b) The absolute values of the rises of the two main transient features around 28.3 eV (VB) and 29.9 eV (CB) are associated with electrons (blue open

circles) and holes (red open squares), respectively, exhibiting a rise time limited by the duration of the VIS-NIR pulse. The measured band shift $\Delta E_{\text{shift}}(\tau)$ (black open diamonds) also follows the carrier excitation within the instrumental response time. The solid lines in (b) are moving averages to guide the eye. In panel (c) the differential absorption of the carrier dynamics directly after valence-to-conduction band excitation, i.e. for positive time delays averaged over $\tau = 8$ to 12 fs (indicated by the black rectangle in (a)), is shown. The dotted black line shows the differential absorption calculated from a TDDFT calculation assuming pulses with a peak intensity of $10^{11}$ W/cm².

**Electron kinetics**

The carrier dynamics $\Delta A_{\text{SB},3d_{5/2}}(E,\tau)$ associated with the electrons (Fig. 4a) are analysed by singular value decomposition (SVD) to separate the underlying signals into different independent contributions of electrons and holes and to trace out their temporal evolution, see Methods section for details. The transient signal can be described by two major components whose spectrum is depicted in Fig. 4b along with the time dependence depicted in Fig. 4c. Calculating a carrier dynamics transient from these two SVD components (Fig. 4d) yields good agreement with the measured carrier dynamics transient (Fig. 4a). The component with larger amplitude and with the spectral distribution peaking near the band edge (red data points in Fig. 4b and c) can be assigned to the thermalized electrons. It exhibits a time dependence that can be fit with a single exponential decay with a time constant of $\tau_{\text{e,recomb}} = (1140 \pm 50)$ fs, which is characteristic of the recombination of the carriers, since this is the time scale when the transient carrier dynamics signal vanishes (Fig. 4a). The error bars are derived from the fitting to the experimental data. A single exponential describes the observed time dependence best and hints at a trap-assisted recombination process[25], which is expected to be dominant in defect rich samples[26]. The second largest component (blue data points in Fig. 4b and c) consists of higher energy contributions and lower energy components with reversed sign. This component (blue data points and fit with dashed black line in Fig. 4c) decays with $\tau_{\text{e,relax}} = (110 \pm 30)$ fs, reversing its sign at the long time limit. This component of the SVD can be interpreted as high-energy electrons (hot carriers) relaxing very rapidly from higher energy states to fill available states at lower energies within this relaxation time.

In addition, single exponential functions were fitted to profiles along the time-delay axis at energies between 29.6 and 30.4 eV (Fig. 4e and f), with time-zero amplitudes and time constants as free parameters (Fig. 4g). The time-zero amplitudes relate to the initial distribution of electrons (green line with shaded error band in Fig. 4g) and show that initially a broad distribution of electrons is excited. The time constants (blue line with shaded error band in Fig. 4g) provide a global view of the energy-specific lifetimes. One observes generally an increasing lifetime of carriers closer to the CB edge. Further, the CB valleys such as $X_1$, $\Delta_1$, $\Gamma_{2'}$ and weakly $L_1$ stand out with significantly longer lifetimes, because the carriers relax to particular valleys and accumulate there.

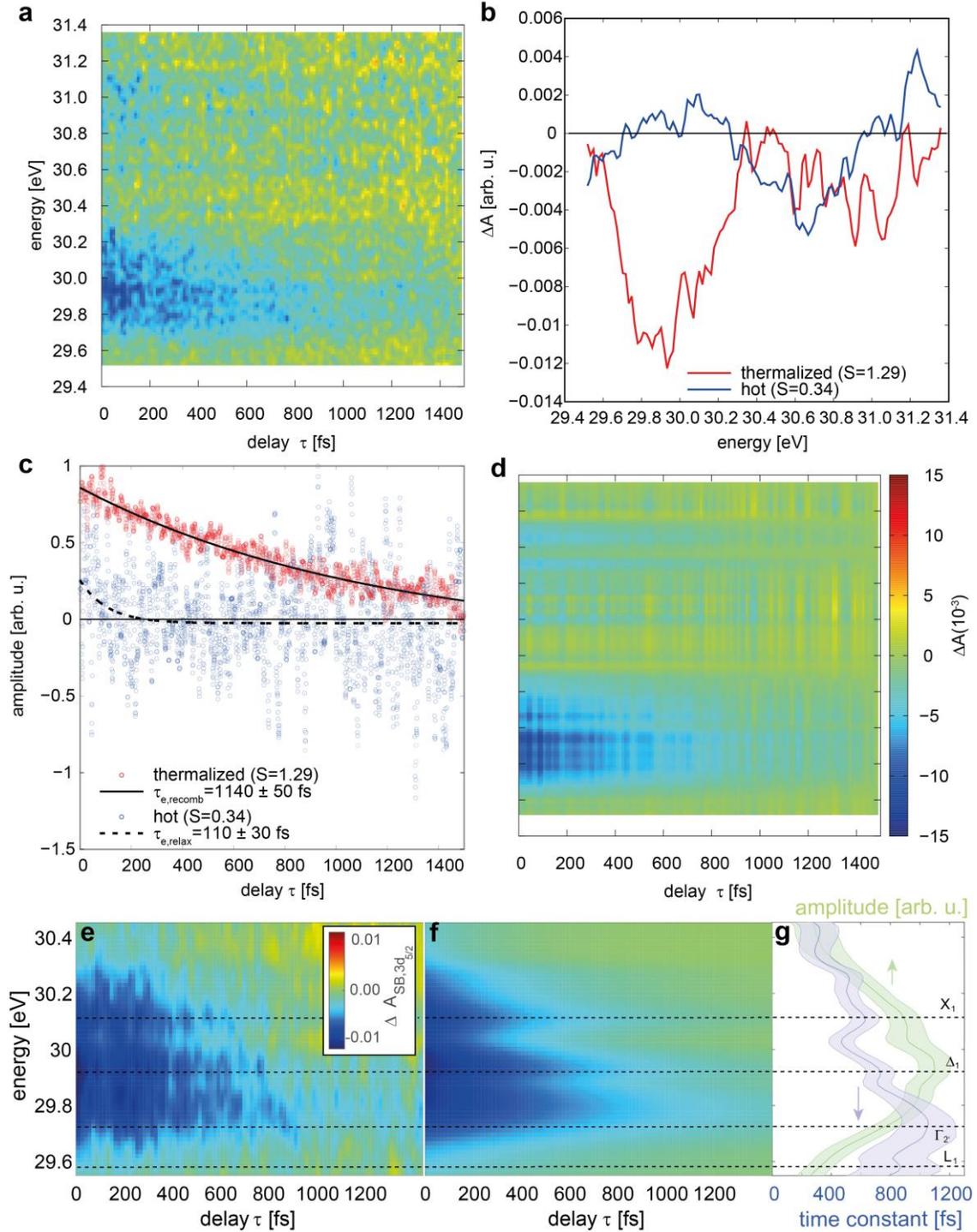

**Figure 4. Electron kinetics in nanocrystalline germanium following ultrafast excitation.** (a) The measured change of absorbance in the carrier dynamics in the spectral region of the CB decomposed into two singular value components. Panels (b) and (c) show the spectrum and the time dynamics of these singular

value components, respectively. (d) Calculating a transient signal from both components and their time-dependence yields good agreement with the measurement. Panels (a) and (d) share the colourbar indicated in (d). The temporal dependence of the stronger component can be best described by a single exponential decay with a time constant of $\tau_{e,\,recomb} = (1140 \pm 50)$ fs associated with the carrier recombination. The weaker component has a time constant of $\tau_{e,\,relax} = (110 \pm 30)$ fs, suggesting a fast relaxation of hot electrons from higher energies to lower energies. (e) For a section of the CB near the band edge, a single exponential decay is fit to each energy of the data, which allows the construction of a map of the dynamics of hot electrons versus energy (f), which is in reasonable agreement with the data. (g) The obtained time constants versus energy (blue line with shaded error band (g)) indicate increased life times of carriers at the CB valleys at several energies, with the longest life times at the low energy valleys, consistent with the SVD analysis. See text discussion for further explanation and interpretation.

**Hole kinetics**

Figure 5a shows the carrier dynamics signal that is associated with the holes, which is also analysed by SVD. The wide positive distribution at time zero spanning from 27.8 to 29.2 eV narrows down and decays as the time delay $\tau$ increases, representing the hot carrier relaxation and recombination of holes. Decomposing the transient absorption signal by SVD reveals two major components. For the largest component, an assignment is made to a thermalized hole distribution (Fig. 5b, red line), whose dynamics show first a rise, $\tau_{split-off,heavy\,hole} = (140 \pm 10)$ fs, after the initial excitation and a subsequent $\tau_{h,recomb} = (1080 \pm 90)$ fs decay (Fig. 5c, red dots with solid black line) that are obtained by fitting a bi-exponential function. The assignment of the faster dynamics with the time constant $\tau_{split-off,heavy\,hole}$ to hole scattering pathways from the split-off to heavy hole band is based on the initial increase of the measured carrier concentrations. This is most clearly seen for time delays of 100-150 fs, where both the measured transients (Fig. 5a) as well as the thermalized component of the SVD (Fig. 5c) maximize. In Fig. 1c it is shown that excitation to the split-off band by the pump pulse, which is of partial 4*s* orbital character (Fig. 1b), is possible. Once carriers scatter from the split-off band to the heavy hole band, which is predominantly of 4*p* orbital character, the carriers become observable to the XUV and increase the carrier dynamics signal. Since the VB is degenerate at the Γ point with predominantly 4*p* orbital character another possible pathway would be hole relaxation within the split-off band towards the Γ point. However, this process has been previously found to be less efficient compared to the split-off-to-heavy-hole band scattering process[14]. For other pathways such as carriers scattering from the heavy or light hole band into the split-off band, a reduction of amplitude, contrary to the observation, would be expected in the carrier dynamics signal.

The second largest component features a positive distribution at larger (hot) hole energies (~28 eV) and a negative distribution near the band edge (Fig. 5b, blue line). The time evolution of this component starts at a positive value and becomes negative at the long-time limit; it is fitted with a single exponential decay (Fig. 5c, blue dots with dashed black line) with a $\tau_{\text{relax}} = (170 \pm 10)$ fs time constant. This indicates that carrier dynamics, albeit a signal proportional to the population distribution, changes sign at the long-time limit. Multiplying the time dynamics and the carrier dynamics component together, it suggests that the second largest component represents a depletion of hot holes and an increase of low-energy holes as the singular vector in time decays from positive to negative value (Fig. 5c, blue dots and dashed black line), representing the hot hole relaxation.

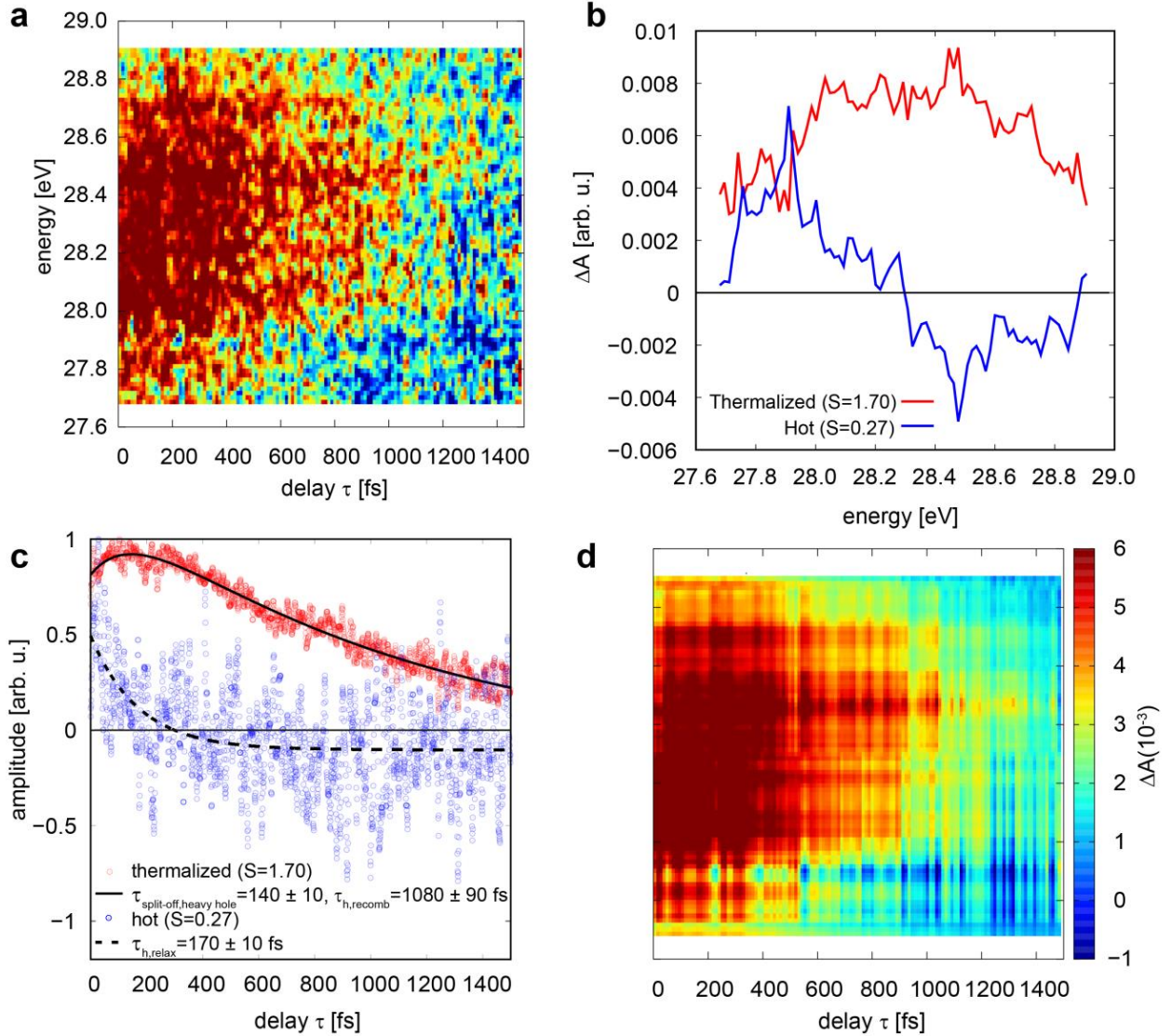

**Figure 5. Hole kinetics in nanocrystalline germanium following ultrafast excitation.** (a) The measured change of absorbance in the spectral region of holes can be decomposed into two singular value components. Panels (b) and (d) show the signal distribution and time dynamics of the two largest singular value components, respectively. (d) Their addition reproduces the observed transient absorption trace. Panels (a) and (d) share the colourbar indicated in (d). The stronger component (red data in (b) and (c)) can be associated with thermalized holes. The growth of this component for $\tau_{\text{split-off, heavy hole}} = (140 \pm 10)$ fs and subsequent decay in $\tau_{h,\text{recomb}} = (1080 \pm 90)$ fs can be understood by holes scattering from the split-off band to the heavy hole band and subsequent recombination with electrons. The weaker component (blue data in (b)

and (c)) can be associated hot holes, which exhibit a fast relaxation to lower hole energies within $\tau_{\text{relax}} = (170 \pm 10)$ fs. See text for further explanations.

## Time-dependent band shifts

The main contribution from the band shift observed here originates from the redshift of the CB, where the absorbance experiences a steep rise (inset Fig. 3c, upper panel). The VB region is attributed to energies where the absorbance is relatively flat, indicating that band shifts in either direction would result in weak contributions on the VB side. This is a feature that XUV transient absorption measurements capture differential quantities, where underlying small shifts, which can be significantly below the actual spectral resolution of the instrument, can cause measurably broad signals if these changes occur in steep sections of the underlying ground state absorbance (cf. Fig. 2e).

In Fig. 6, the heat-induced band shift has been subtracted in order to focus on the band shift following photoexcitation using different intensities. The carrier-induced band shift scales with the third root of the number of excited carriers $\Delta E_{\text{gap}} \sim N_e^{\frac{1}{3}}$ (see Sec. S6 in Supplementary Information) and is thus time-dependent as the carrier density changes[21], e.g. by recombination. In the experiment, with increasing intensity and thus increased initial carrier density $N_e$ in the CB, the observed redshift increases. In Fig. 6 the time-dependent energy shift at one excitation intensity (black line in Fig. 6) is compared to $\Delta E_{\text{gap}}(\tau) \sim N_e(\tau)^{\frac{1}{3}}$ (red dotted line in Fig. 6) using a carrier density decay that follows an exponential with a time constant of 1.1 ps with increasing delay $\tau$. The excellent agreement suggests that the observed band shift is predominantly caused by a carrier-induced dynamic redshift of the CB.

The excited carrier densities range from $0.4 \times 10^{21} \frac{1}{\text{cm}^3}$ to $1.2 \times 10^{21} \frac{1}{\text{cm}^3}$ according to the TDDFT calculations using the pulse parameters of the experiment. In the inset of Fig. 6 the measured redshifts directly after excitation for the calculated carrier densities are compared to $\Delta E_{\text{gap}}/2$ (red dashed line, inset Fig. 6) to account for the fact that here only the redshift of the CB is observed. Although the order of magnitude agrees, small differences between experiment and calculation suggest that besides a pure carrier-induced band shift, additional effects such as core-level shifts and phonon renormalization may play a role. This, however, does not limit the extraction of the carrier dynamics contribution. An extended discussion can be found in Sec. S6 in the Supplementary Information.

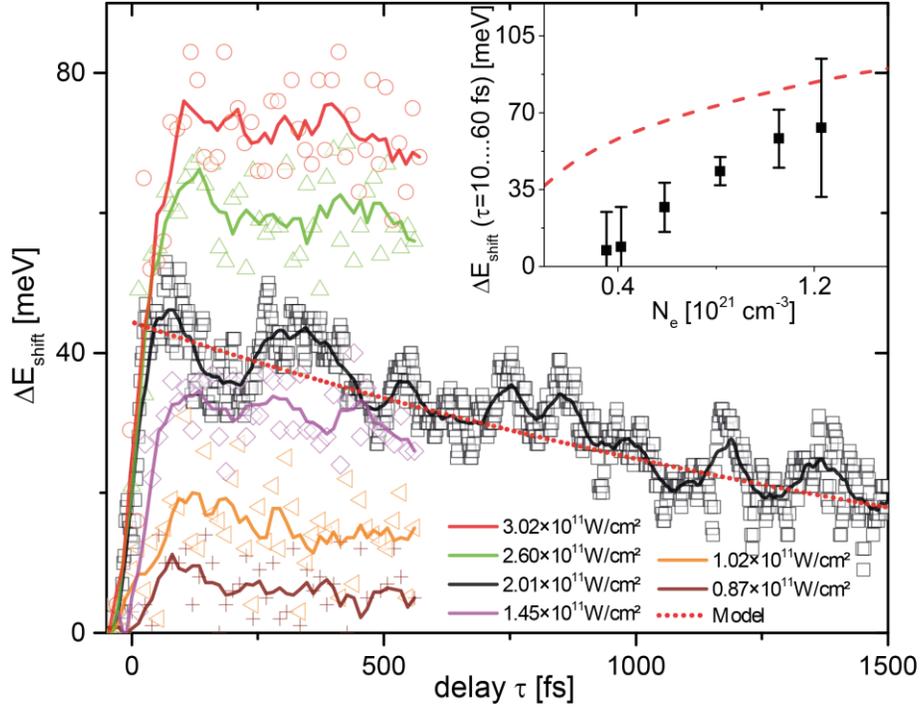

**Figure 6. Band shift depending on time delay and intensity.** Here the band shift $\Delta E_{\text{shift}}(\tau)$ for different excitation intensities is depicted. The temporal behavior suggests that the observed shift is predominantly due to a redshift of the CB due to carrier-induced band shift $\Delta E_{\text{gap}}(\tau) \sim N_e(\tau)^{\frac{1}{3}}$, assuming a single exponential carrier decay with a time constant of 1.1 ps (red dotted line). In the inset the measured initial shifts (black squares) for different initial carrier densities is compared to an analytic calculation of the band shift. The calculated redshift of the CB (red dashed line, inset) for different carrier densities is slightly larger than the measured band shifts, which suggests that there are additional contributions by core-level shifts and phonon renormalization.

## Discussion

In the present experiment, electron and hole dynamics can be studied simultaneously and with high time and energy resolution. The energetic separation of the electron and hole signals is of great value since band-related features and carrier-specific dynamics can be unambiguously assigned (Fig. 3a). The decoupling of the pump and probe spectral regions allows new possibilities to excite samples for instance with ultrashort broadband laser pulses as in the present experiment to study ultrafast excitation dynamics and electronic relaxation processes of both carriers, or with longer monochromatic pulses for valley-specific excitation *independent* of the XUV probe pulse. The possibility for frequency-converting the ultrafast pump pulses will further extend the accessible materials. Converting the ultrashort pump pulse into the ultraviolet-to-blue spectral region[27] will allow the investigation of wide-bandgap materials. ATAS provides not just the necessary temporal resolution to, in principle, capture sub-cycle carrier dynamics of both carrier species[28], but also the necessary continuous XUV bandwidth to capture the more than 4 eV wide dynamical features (Fig. 3a) in parallel.

Discussing the electron dynamics, the assignment of the CB valleys in Fig. 4e is essential. Using energy-resolved line profiles and fitting single exponential decays at each energy for near-edge CB states, energy-resolved carrier life times are extracted. The initial amplitudes of the exponential fits (green line with shaded error bars in Fig. 4g) suggests that initially a broad distribution of electrons is excited. Following excitation, the carriers undergo intervalley and intravalley scattering with phonons, effectively populating all bands within the energy range and relaxing the electrons to respective valleys. The measured time constant versus energy (blue line with shaded error bar in Fig. 4g) supports this intuitive picture. There, the time constant for higher-lying hot carrier states is the shortest, continually increasing towards the band edge. At the valleys, the measured time constants peak due to the carriers accumulating (see e.g. energy assigned to $X_1$ and $\Delta_1$ valleys in Fig. 4g). Hence, although the experiment averages over all *k* points in the Brillouin zone, the results suggest that characteristics of specific valleys can be extracted from an ATAS experiment due to the high DOS at the valleys, provided the involved valleys have sufficient energy spacing. In ATAS at the germanium $M_{4,5}$-edge this is the case e.g. for the $X_1$ valley versus $\Gamma_{2'}$, but not for the $L_1$ and $\Gamma_{2'}$. Here, the achievable resolution is mainly limited by the core-hole lifetime broadening, which constrains the energy resolution to $\delta E \sim 0.24$ eV (ref. 20). For nanocrystalline samples an additional broadening due to averaging over different crystal orientations is expected[29]. In the experiment, the slow time constants for decay in the $\Gamma_{2'}$ and $L_1$ valleys are on the order of ~1.1 ps, which is consistent with the carrier decay time in the SVD analysis. The SVD components (Fig. 4b) can be understood as an ensemble average and the measured ~100 fs relaxation time in the second SVD component corresponds to an average relaxation time that also includes electrons at higher energies, where the signal strength prohibits meaningful line profile measurements.

In the hole kinetics, the initial rise of the main component of the hole transient (Fig. 5c) has been assigned to holes scattering from the split-off band into the heavy hole band based on comparison of orbital characters in the VB (Fig. 1b) and only one intervalley scattering pathway that allows for a signal increase. This assignment is supported by comparison to hole relaxation times measured in germanium by optical pump-probe techniques. Woerner *et al.* reported intervalley scattering times of holes from the split-off band assisted by phonons to the heavy-hole band in the $\Gamma$ valley to be within their instrument response time of 250 fs and estimated a lifetime of ~100 fs by calculations[14]. This value corresponds well to the (140±10) fs rise of the main component obtained in the present experiment. Although Woerner *et al.* excited specifically to the split-off band from the heavy hole band near the $\Gamma$ point, their ~0.5 eV photon energy results in comparable hole energies as in the present experiment, considering that a broadband pulse with ~1.65 eV central photon energy excites across the band gap ($\geq 0.8$ eV). Due to the higher effective mass of the heavy holes, the scattering process towards the heavy hole band is expected to be dominant and the model calculations and experimental results in their paper suggest the interband scattering from the split-off band is more effective than interband relaxation. The long time decay of the hole signal on the order of ~1.1 ps indicates the recombination of the holes with the electrons, which within the error bars have the same decay constant for the observed time window and accessible signal-to-noise ratio.

In the present experiment, the electrons and holes exhibit a similar lifetime, i.e. the decay constants are ~1.1 ps for both. These lifetimes appear to be more than two orders of magnitude shorter compared to single crystal germanium[30]. A question that arises is if whether the electrons and holes recombine or if they relax to states that are invisible for the experiment, e.g. states of *s* orbital character. The *in situ* measurement of the band shift suggests an interpretation of carrier recombination within the observed time scale, because the band shift over time (cf. Fig. 6) also follows a behaviour that indicates a single exponential decay of the free carriers in the CB as is observed in the kinetic analysis of the carrier dynamics in the CB, with agreement of the time constants. The longest lifetime occurs in the lowest bands, compared to a shorter lifetime at higher energies, supporting that the carriers have first thermalized and subsequently recombine. A possible recombination mechanism that would support the observed fast decay rate suggests the involvement of intermediate trap states and a Shockley-Read-Hall (SRH) recombination mechanism[26,31,32]. This can be understood based on the nanocrystalline samples. The nanocrystalline thin films have a grain size of ~11 nm. Further it is known that randomly oriented non-passivated nanocrystals exhibit a large number of trap states at the interfaces of the nanocrystallites. These trap states can act as recombination centers[33]. The recombination times are expected to be significantly accelerated compared to single-crystalline samples, since assisted by phonon-scattering the carriers can quickly scatter into these trap states and subsequently recombine.

The observed time scales are in good agreement with visible transient reflectivity measurements on germanium nanorods, where carrier life times of ~6 ps were measured for nanorods having a diameter of 18 nm (ref. 34), considering that grown nanorods in general are expected to have a lower number of trap states compared to annealed nanocrystalline thin-films. Further, the smaller domain size in the present experiment and three-dimensional versus two-dimensional trap state arrangement locations can explain the comparatively shorter measured life times.

In summary, this paper presents the study of ultrafast carrier relaxation in germanium using NIR-XUV pump-probe spectroscopy with few-femtosecond time resolution. We demonstrate that transient absorption in the XUV yields clear, spectrally resolved signatures for electrons and holes, which allows tracking of electrons and holes independently after a broadband (500-1000 nm) optical excitation. An analysis is established that allows the decomposition of the transient absorption data into contributions from state blocking, band shifts and excited state broadening. It is shown that after retrieving the state blocking contribution, a separation of the observed signal into the contribution from a single spin-orbit state at the germanium $M_{4,5}$-edge is possible in order to retrieve the carrier dynamics. Comparison with first principles calculations supports the assignment of electron and holes in the retrieved carrier dynamics transient signal. Due to the high temporal and energy resolution this is an ideal tool for disentangling the contributions of relaxation and recombination processes. Clear signatures of hot carrier relaxation for both electrons and holes in germanium occurring below 200 fs is observed, which has been so far inaccessible through narrowband, pure optical/NIR pump-probe techniques. In addition, the presented experiment suggests that the retrieved time-dependent band shift can be decomposed into a heat-induced and predominantly a carrier-induced redshift of the conduction band. The heat-induced redshift allows for *in situ* characterization of the sample temperature. The temporal behaviour of the carrier-induced redshift supports the observed fast ~1.1 ps carrier recombination of electrons and holes observed in the carrier dynamics. Moreover, scattering and relaxation rates can be measured in the time domain and by the additional high energy resolution quantitative data can be retrieved for different parts of the band structure. This technique will be an invaluable tool for research on more complex semiconductor materials such as strained layers, heteroatomic, ternary, quaternary systems, junctions, two-dimensional material compounds and quantum confined systems, which are becoming increasingly important for applications such as solar energy production and highly efficient computer processors.

## Methods

*Experimental details and data analysis*

NIR pulses with 5 fs pulse duration are used to optically excite germanium thin films. The nanocrystalline germanium films, fabricated by electron beam deposition and subsequent annealing, are 100 nm thick and supported by a 30 nm thick silicon nitride membrane. The crystallites have an average size of 11 nm, see Supplementary Information Sec. S7 for extended characterization. A spectrally continuous XUV pulse spanning the germanium $M_{4,5}$-edge is generated by high harmonic generation in xenon assisted by PASSAGE gating[35]. The XUV pulses are used to probe the transient population in the VB and CB in a transmission geometry and are subsequently resolved by a flat-field spectrometer. By changing the time delay $\tau$ between the pulses the transient changes are measured. A shutter periodically blocks the VIS-NIR pump beam to obtain a differential absorption signal. The data set shown in main text Fig. 2 and Fig. 3 (panel a, b, c and black line in Fig. 6) and analysed in Fig. 4 & 5 consists of five averages and was scanned with 0.6 fs time steps around time zero and with 3.3 fs time steps out to 1.5 ps. The additional intensities shown in Fig. 6 consist of 50 averages each with 26 fs time steps on an interval of $\tau$=[-60, +550 fs]. To compensate for time delay drifts in parallel to the germanium sample the differential absorption at the argon $3s3p^64p$ autoionizing state is measured[36]. The instrumental response time has an upper bound estimated to ~7 fs by comparing these gas transient absorption measurements. At the same time the spectrometer resolution can be estimated to $\delta E \sim 0.07$ eV from known line widths in gas absorption spectra. In the post-experimental data analysis first the measured differential absorption data $\Delta A_{meas}(E, \tau)$ data are decomposed into contributions from state blocking, band shifts and excited state broadening. The high quality static absorbance used for decomposing these contributions was measured by averaging one hundred transmission spectra alternating between the sample and an empty silicon nitride membrane under otherwise identical conditions to those for the time-resolved measurements. The state blocking signal is subsequently separated into a single spin-orbit split state by dividing out a constant phase factor in the Fourier domain originating from the spacing ($\Delta E_{SO,Ge} = 0.58$ eV, ref. 20) and degeneracy of the core-levels ($g_{3d_{5/2}} = 6$ and $g_{3d_{3/2}} = 4$) exhibiting spin-orbit splitting and subsequent back transformation:

$$\Delta A(E) = \int \frac{\Delta \tilde{A}_m(\eta) \, e^{i\eta E}}{6 + 4e^{-0.58 i\eta}} \, d\eta. \tag{1}$$

The retrieval constitutes the signal probed from the *3d₅/₂* state. Further details are given in the Supplementary Information Sec. S4.

*Excited carrier-density*

The number of excited electrons has been estimated in two ways. Firstly, from the fluence $F$ of the laser pulse with central frequency $\nu$ and an absorption coefficient $\alpha$ of a sample with thickness $d$, one can estimate for a one-photon transition, where one absorbed photon excites one electron, that $N_e \cong \frac{F}{h\nu d}[1 - \exp(-\alpha d)]$. For the presented experiment ($I = 2 \times 10^{11} \frac{W}{cm^2}$, $F = 12.4 \times 10^{-3} \frac{J}{cm^2}$, $d = 100$ nm, $\alpha = 49 \times 10^3 \frac{1}{cm}$, $\lambda_0 = 760$ nm) one gets $N_e \cong 8 \times 10^{20} \frac{1}{cm^3}$. Secondly, the performed TDDFT calculations using $k$-dependent excitation and the spectrum of the experimental laser pulse an excitation fraction of 0.3% was calculated for a laser intensity of $2 \times 10^{11} \frac{W}{cm^2}$. Using that a unit cell in germanium has a volume of $V_0 = 4.527 \times 10^{-23} cm^3$ and there are two atoms or 8 valence electrons in this volume, one gets a valence electron density of $d_0 = 1.76 \times 10^{23} \frac{1}{cm^3}$. Hence, for the TDDFT-based calculation one gets $N_e \cong 5 \times 10^{20} \frac{1}{cm^3}$.

*Heat-induced band shift*

In the time-dependent transient absorption experiments a transient signal at negative time delay, i.e. NIR pump arrives after XUV probe, was consistently observed when measuring first the unexcited (cold) sample transmission followed by the excited (hot) sample spectrum in order to obtain a differential absorption spectrum. A comparison of the absorption change in a heated Ge film to this spectral feature, suggests it is related to an increased temperature of the germanium thin film after optical excitation causing a heat-induced band shift[37]. The absorption edge redshifts causing a positive differential signal. By resolving the underlying shift due to thermal expansion of the lattice $\Delta E_{ren,therm}$, a temperature can be assigned to the thin film. For example, for an average pump power of $P_{avg} = 0.39$ mW the measured shift of the edge is $\Delta E_{ren,therm} = 72.8$ meV suggesting a temperature of $T = 475$ K, which is confirmed by heat diffusion calculations using a finite elements method (see Sec. S5 in Supplementary Information).

*First principles TDDFT and XAS calculations for germanium*

To support the assignment of spectroscopic features, X-ray absorption spectroscopy calculations based on the eXcited-electron Core-Hole (XCH)[38] approach were combined with first principles time-dependent density functional theory (TDDFT)[39,40]. This first-principles treatment of the electric field dynamics described using real-time TDDFT[39,40] includes effects beyond the linear response of the material, and similar theoretical treatments reproduce electron tunnelling in silicon[7] and the dielectric breakdown in insulators[41]. The exciting VIS-NIR pump laser field models a few fs, 780 nm pulse with a peak intensity of $1\times10^{11}$ W/cm$^2$ (see Fig. S7 in Supplementary Information).

The excited-state electron and hole occupations generated by interaction with the pump laser pulse are obtained from the real-time TDDFT calculation in the form of a single-particle density matrix (DM). This one-electron DM is subsequently projected onto an eigenbasis of the core-excited XCH Hamiltonian, along the lines previously described[7,42,43], to estimate the $M_{4,5}$-edge XVU absorption in the presence of excited carriers. The difference between this excited state spectrum and the corresponding ground state spectrum in the absence of electron-hole pairs, is compared to the *3d$_{5/2}$* core-level differential absorption retrieved from experiment at time delays directly after excitation, shown in Figure 3c. Inclusion of the core-hole excitonic effects, at the level of the XCH approach, in the core-level absorption calculation was necessary to match the magnitude of the negative feature at 30 eV. In the first principles calculations the excited and ground state spectra are treated only for a single spin-orbit split state and shifts, such as core-level shifts or band shifts, are not included. We note that at the level of the adiabatic TDDFT and XCH theories employed here we expect state-blocking effects in the excited state $M_{4,5}$-edge absorption to be captured but many-body effects such as excited-state band-gap shifts and carrier life-time modulation are not described. Therefore, the comparison with experiment is invoked primarily in the context of state-blocking where good agreement is observed (Fig. 3c). Further details are given in Sec. S10 in the Supplementary Information.

*Singular value decomposition*

The carrier dynamics can be decomposed into several carrier distributions with independent relaxation dynamics. Under such formulation, the transient absorption signal $\Delta A(E, t)$, with $E$ as photon energy and $t$ the delay time, can be decomposed into

$$\Delta A(E, t) = \sum_n s_n u_n(E) v_n(t), \tag{2}$$

where $u_n$ is the $n^{th}$ component of transient absorption which corresponds to a distinct carrier distribution and $v_n$ its accompanying relaxation dynamics. $s_n$ is the singular value of the $n^{th}$ component which signifies the importance of the component on the overall dynamics.

*Germanium band structure calculation*

The band structure of germanium (Fig. 1b) is calculated by density functional theory (DFT) within the Vienna ab-initio simulation package (VASP)[44,45], using a generalized gradient approximation and Hubbard (GGA+U) approach. The computational method and parameters are detailed in Ref. 46.

**Acknowledgements**

The initial instrument development and experimental work was supported by the Office of Assistant Secretary of Defense for Research and Engineering through a National Security Science and Engineering Faculty Fellowship (NSSEFF) and W. M. Keck Foundation. M. Z. acknowledges support by the Army Research Office (ARO) (WN911NF-14-1-0383). H.-T. C. and L. J. B. acknowledge support by the Air Force Office of Scientific Research (AFOSR) (FA9550-15-1-0037). Additional funding for L. J. B. was provided by NSSEFF. Additional funding for C. J. K. was provided by the Defense Advanced Research Projects Agency PULSE program through grant W31P4Q-13-1-0017. J. S. P. and A. G. acknowledge support by NSSEFF.  S. K. C. acknowledges a postdoctoral fellowship through the Office of Energy Efficiency and Renewable Energy of the Department of Energy. P. M. K. acknowledges support from the Swiss National Science Foundation (P2EZP2_165252).  M. Z. acknowledges support from the Humboldt Foundation. The Department of Energy under contract DE-AC03-76SF00098 is acknowledged for additional experimental equipment. C.D.P. and D.P. performed work at the Molecular Foundry, supported by the Office of Science, Office of Basic Energy Sciences, of the U.S. Department of Energy under

## Author contributions

M. Z. and H.-T. C. conducted the experiments used in this manuscript, established the models and methods and subsequently applied these for evaluating the data. L. J. B. and A. G. conceived the experiment and performed preliminary experiments. C. D. P. and D. P. provided theory support and performed the TDDFT and XAS calculations. M. H. O. performed the sample characterization. D. M. N. and S. R. L. supervised the work. M. Z. wrote the majority of the manuscript with inputs from H.-T. C, P. M. K. and S. K. C.. All authors discussed the results and contributed to the manuscript.

**The authors declare no competing financial interests.**

**The manuscript is accompanied by supplementary material.**

*Supplementary Material for*

# Direct and Simultaneous Observation of Ultrafast Electron and Hole Dynamics in Germanium

Michael Zürch[1,†,*], Hung-Tzu Chang[1,†], Lauren J. Borja[1,†], Peter M. Kraus[1], Scott K. Cushing[1], Andrey Gandman[1,‡], Christopher J. Kaplan[1], Myoung Hwan Oh[1,2], James S. Prell[1,§], David Prendergast[3], Chaitanya D. Pemmaraju[3,4], Daniel M. Neumark[1,5,*], Stephen R. Leone[1,5,6,*]

[1] Department of Chemistry, University of California, Berkeley, CA 94720, USA

[2] Materials Sciences Division, Lawrence Berkeley National Laboratory, Berkeley, CA 94720, USA

[3] The Molecular Foundry, Lawrence Berkeley National Laboratory, Berkeley, CA 94720, USA

[4] Stanford Institute for Materials & Energy Sciences, SLAC National Accelerator Laboratory, Menlo Park, CA 94025, USA

[5] Chemical Sciences Division, Lawrence Berkeley National Laboratory, Berkeley, CA 94720, USA

[6] Department of Physics, University of California, Berkeley, CA 94720, USA

[‡] Present address: Solid State Institute, Technion – Israel Institute of Technology, Haifa, 32000, Israel

[§] Department of Chemistry and Biochemistry, University of Oregon, Eugene, OR 97403, USA

[*] Correspondence should be addressed to: mwz@berkeley.edu (M. Z.); dneumark@berkeley.edu (D. M. N.); srl@berkeley.edu (S. R. L.)

[†] These authors contributed equally to this work.

**S1 Experimental Methods:**

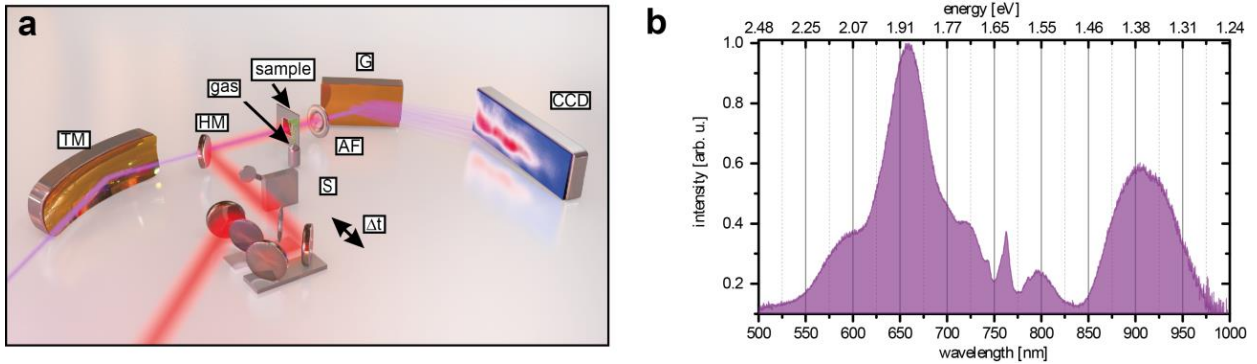

*Figure S1:* Detailed experimental setup. (a) The XUV probe pulse is refocused by a toroidal mirror (TM) onto the sample. A grating (G) disperses the XUV which is measured with a CCD. The VIS-NIR pump pulse (spectrum shown in (b)) is time delayed with respect to the XUV pulse and focused onto the sample. Both beams are collinearly recombined by means of a hole mirror (HM). An aluminium filter (AF) blocks the VIS-NIR light before the spectrometer. A shutter (S) in the pump arm periodically blocks the pump light for differential absorption measurements.

In Fig. S1a the XUV transient absorption apparatus is illustrated. Carrier envelope phase (CEP) stabilized pulses centred at 790 nm, 25 fs in duration are produced using a commercial Ti:sapphire chirped pulse amplifier at a repetition rate of 1 kHz. After spectrally broadening these pulses in a neon-filled hollow-core fibre, these pulses are further compressed via chirped mirrors (PC70, Ultrafast Innovations) to ~4 fs pulse duration. A chopper before the beam is split by a 90/10 beam splitter into the probe and pump arm, respectively, reduces the repetition rate to 100 Hz to reduce the average power, minimizing damage of the solid-state sample from heating by successive pulses. Fused silica (FS) wedges in both the pump and probe arm allow for fine-tuning the dispersion. The proper compression of the pulses is confirmed by the dispersion scan technique[1].

The probe arm proceeds through a 75 μm thick quartz plate followed by a quarter-wave plate for PASSAGE gating[2] and is subsequently focused inside a 1 mm gas cell filled with 28 torr of xenon gas to generate high harmonics. The residual VIS-NIR in the probe arm is blocked by a 200 nm thick aluminium filter. A gold-coated toroidal mirror focuses the XUV by direct imaging the source onto either a solid-state thin film or another 1 mm gas cell for characterization measurements. The focal spot size of the XUV radiation is estimated to have a diameter of ~50 µm full-width half maximum (FWHM) by a knife-edge scan. The XUV light that is transmitted through the sample is measured by a flat-field spectrometer. The spectrometer was calibrated using argon and neon autoionizing states between 26-48 eV (refs. 3,4), and the spectral resolution was determined to be ~70 meV. The typical XUV continuum spans 25 to 40 eV (see main text Fig. 1d) and exhibits CEP effects if the CE is scanned, indicating the attosecond nature of the pulse.

The VIS-NIR pump pulse propagates alongside the vacuum apparatus and is collinearly recombined with the XUV beam by means of an annular mirror after the toroidal mirror. The VIS-NIR pump pulse covers photon energies from 1.2 to 2.5 eV (Fig. S1b) and can be blocked by a shutter in order to measure the absorption change with and without the pump pulse. The VIS-NIR pump arm is focused onto the target with a beam diameter of ~100 µm (FWHM). A piezo-driven delay stage in the pump arm allows for precise control of the relative time delay between the pump and probe pulses. An additional 200 nm thick aluminium filter after the sample blocks the pump arm before the spectrometer. An iris is used for regulating the pump arm to specified intensities.

To correct for thermally induced time delay drifts, the experiment alternates between measuring pump off and pump on transmission spectra on the germanium sample and an argon gas cell for each individual time step. The germanium sample was raster scanned to randomly chosen positions on a regular grid of points spaced by 150 microns in both dimensions over an area of ~2.5 x 2.5 mm² to ensure averaging out possible local inhomogeneities of the thin film. Typical integration times are on the order of one second to optimally use the dynamic range of the detector. However, for signal levels in the $\Delta A <$ 0.05 level, averaging several data sets is required for improving the signal-to-noise ratio. The typical number of averages range from 5 to 50, which takes between 4 and 24 hours for a complete experiment at all time delays, depending on the number of time steps. Hence, time delay drifts in the experimental apparatus must be corrected for. Here, it was chosen to scan all time delays by alternating between the germanium sample and the argon gas cell and repeat the full scan for a number of averages. The $3s3p^64p$ autoionizing state in argon produces a sufficiently strong transient signal[5] in each single scan. Fitting a Boltzmann sigmoidal function to the normalized rise of this feature

$$\Delta A_{Ar,3s3p^64p}(t) = \frac{1}{1 + \exp[(t - t_0)/dt]}, \tag{1}$$

where $t_0$ and $dt$ are the time of the signal reaching the 50% level and the slope at that point, respectively, allows determining a common time base. Although this approach does not reflect the complexity of field-induced line shape changes,[6] nor does it produce an absolute time zero[7], it allows an effective determination of the time delay drifts between different scans *in situ* on target and without changing the experimental scheme. Further, monitoring *dt* allows to confirm constant pump pulse characteristics over the course of the experiment, assuming that slight differences in pump intensity would modify the ponderomotively driven line shape changes significantly. For the experiments presented here, typical time delay drifts between successive scans were on the order of one femtosecond, which were compensated in post-experimental analysis by interpolating all data sets onto a common time delay axis.

## S2 Calculation of *k*-dependent excitation cross section

The measured change of absorbance $\Delta A_{\text{meas}}$ in the XUV is proportional to the population of holes, $N_h$, and electrons, $N_e$, in the valence and conduction bands. The population of holes is proportional to the oscillator strength $\Omega(\vec{k})$ and an excitation probability $P_{\text{abs}}$, assuming a uniform oscillator strength of unity and therefore determined by the spectrum of the VIS-NIR pump pulse $I(\omega)$ and the energy gap between valence and conduction bands, $\omega_g(\vec{k})$:

$$N_h(\vec{k}) \sim \Omega(\vec{k}) P_{\text{abs}}, \tag{2}$$

and

$$P_{\text{abs}}(\vec{k}) = I(\omega) \sum_{\{i,j\}} \delta(\omega - \omega_g^{ij}(\vec{k})), \tag{3}$$

where $i$ and $j$ are band indices for valence and conduction bands, respectively. Figure S2 shows the excitation probability at different *k* points in the heavy-hole, light-hole, and split-off bands. The calculation includes excitation to the four low-lying conduction bands and assumes a constant transition dipole element.

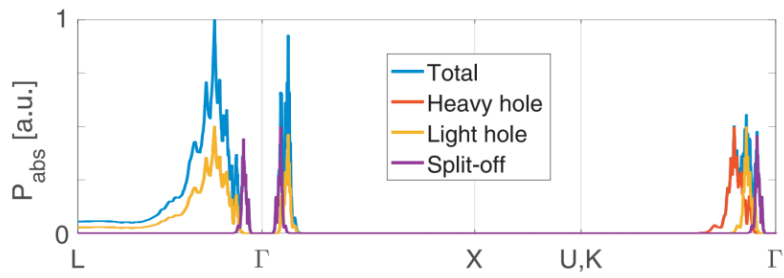

*Figure S2: Allowed hole excitations at different k points in the heavy-hole, light-hole, split-off band, and the total contribution from the three for the VIS-NIR pulse used in the experiment for germanium. Note that between the L and X points the heavy and light hole band are degenerate, i.e. the yellow line covers the red line.*

## S3 Extraction of state blocking, band shifts and broadening in time-dependent differential transient absorption data at the Ge $M_{4,5}$-edge

The assignment of features in a measured raw $\Delta A$ trace is hampered by several physical effects that overlap spectrally. The change of absorbance ($\Delta A$) includes contributions from electronic state blocking from carrier excitation, spectral shift of the M-edge due to core-level shifts and band shifts from excited electron-hole plasma and phonons, and spectral broadening of the excited state spectrum. To retrieve the carrier dynamics from the transient absorption signal, we decompose the measured change of absorbance, $\Delta A_{\text{meas}}(E, \tau)$, (see main text Fig. 2a) into contributions from electronic state blocking $\Delta A_{\text{SB}}$, spectral shift $\Delta A_{\text{shift}}(E, \Delta E_{\text{shift}}(\tau))$ of the static absorption spectrum ($A_{\text{meas,static}}(E)$, main text Fig. 2b) by energy $\Delta E_{\text{shift}}$, and spectral broadening $\Delta A_{\text{broad,n}}(E, \sigma(\tau))$. The broadening is modeled by a Gaussian broadening of the static absorption spectrum by width $\sigma$. Regarding the broadening, assuming a Gaussian functional for broadening relates to phonon-induced broadening while a Lorentzian functional relates to electronic-induced broadening. Since for the presented experiment the component due to broadening in general was found to be negligible, the functional due to broadening was not further evaluated, although it was retained in the analysis because for other materials or parameters it could become important. We fit the measured change of absorption at each time delay with weighted numerical minimization of the following function over the energy range $E = [27.5, 30.5]$ eV:

$$\epsilon_n = \left|\Delta A_{\text{meas}}(E, \tau) - \Delta A_{\text{broad,n}}(E, \sigma(\tau)) - \Delta A_{\text{shift,n}}(E, \Delta E_{\text{shift}}(\tau)) - \Delta A_{\text{SB,n-1}}(E, \tau)\right|, \quad (4)$$

where $n$ denotes the step in the iterative minimization procedure, and

$$\Delta A_{\text{broad,n}}(E, \sigma(\tau)) = A_{\text{meas,static}}(E) \star \exp(-E^2/[2\sigma(\tau)^2]) - A_{\text{meas,static}}(E) \quad (5)$$

$$\Delta A_{\text{shift,n}}(E, \Delta E_{\text{shift}}(\tau)) = A_{\text{meas,static}}(E - \Delta E_{\text{shift}}(\tau)) - A_{\text{meas,static}}(E), \quad (6)$$

where the star operator denotes a convolution. The initial guess for the transient absorption induced by state blocking $\Delta A_{\text{SB,0}}(E, \tau)$ is modelled as follows.

As a first guess to seed the iterative algorithm, we assume the initial carrier distribution resembles Gaussians and their spin-orbit contributions to the transient absorption signal can be written as

$$\Delta A_{SB,0}(E,\tau) = \frac{|\Delta A_{meas}(28.3\ eV, \tau)|}{\max|\Delta A_{meas}(28.3\ eV, \tau)|} \left( 6\ \exp\left\{-\left[\left(E - E_F - \frac{E_c}{2}\right)/\Delta E_{BW}\right]^2\right\} \right.$$
$$+ 4\ \exp\left\{-\left[\left(E - E_F - \frac{E_c}{2} - \Delta E_{SO}\right)/\Delta E_{BW}\right]^2\right\}$$
$$+ 6\ \exp\left\{-\left[\left(E - E_F + \frac{E_c}{2}\right)/\Delta E_{BW}\right]^2\right\}$$
$$\left. + 4\ \exp\left\{-\left[\left(E - E_F + \frac{E_c}{2} - \Delta E_{SO}\right)/\Delta E_{BW}\right]^2\right\} \right). \tag{7}$$

Electrons and holes contribute two Gaussians each in the above expression due to spin-orbit splitting $\Delta E_{SO}$ in the *3d* core level. The pre-factors 6 and 4 originate from the degeneracies of the 3*d* core levels. It is important to mention here, that this step of generating a seed for the state blocking contribution in the iterative algorithm and taking spin-orbit splitting into account is essential, since we choose to decompose the components by using the measured raw transient absorption signal and the static absorbance, both of which have contributions by spin-orbit splitting in germanium. In Sec. S4 of this document we indicate how in a step subsequent to retrieving the state blocking component by the iterative procedure outlined here the spin-orbit contributions are separated. Assuming a Gaussian shape of the state blocking here, as first approximation, is chosen as a reasonable compromise between the energy resolution of the instrument and the initial distribution expected from the spectrum of the excitation pulse, which would translate into a Fermi-Dirac type distribution after thermalization. Here, we assign the Fermi energy $E_F = 29.2$ eV, the average optical excitation energy $E_c = 1.65$ eV, and a bandwidth of excitation $\Delta E_{BW} = 1$ eV according to the experimental parameters. The state-blocking-contributed transient absorption signal is refined and retrieved at every iteration step $n$, using

$$\Delta A_{SB,n}(E,\tau) = \left[2\left(\Delta A_{SB,n-1}(E,\tau)\right) + \left(\Delta A_{meas}(E,\tau) - \Delta A_{broad,n}(E,\sigma(\tau)) - \Delta A_{shift,n}(E,\Delta E_{shift}(\tau))\right)\right]/3 \qquad (8)$$

where the weighting factors are chosen to aid smooth convergence. Note that during the iterations no particular treatment of spin-orbit splitting is necessary, since all components considered and input into the procedure have contributions by the spin-orbit split 3*d* core level in germanium. In this way the iterative procedure works as closely as possible on the experimentally measured data, i.e. the transient absorption signal and the static absorbance, which seeks to improve numerical stability of the procedure.

After five iterations a stable state blocking signal is retrieved (main text Fig. 2c) along with a delay dependent broadening $\sigma(\tau)$ and $\Delta E_{shift}(\tau)$. In main text Fig. 2d and e the resulting contributions for $\Delta A_{broad}(E,\sigma(\tau))$ and $\Delta A_{shift}(E,\Delta E_{shift}(\tau))$ are shown. Figure S3 depicts the first iteration of the decomposition. Already with the first guess for the state blocking based on measured experimental parameters, reasonable agreement between the measured $\Delta A_{meas}(E,\tau)$ trace and a calculated trace based on $\Delta E_{shift}(\tau)$ and $\sigma(\tau)$ after the first iteration is achieved (Fig. S3a & b). The major contribution besides state blocking comes from an underlying redshift (Fig. S3c). The decomposition suggests that broadening appears to play a minor role and contributes less than 5% to the signal. From the retrieved state blocking, i.e. when the iterative procedure has converged, one can subsequently retrieve the spin-orbit separated state blocking by applying the method outlined in Section S4 (main text Fig. 2f); the result can then be further analysed for the underlying carrier dynamics. It is important to note that this model only assumes a linear shift for all energies. As can be also seen for the analysis of the heat-induced band shift (Fig. S5b), the agreement of the experiment and model is reasonable below ~31 eV. At higher energies the disagreement can be explained by different critical points in the band structure, which shift by different amounts compared to the main contributions at the bottom of the CB. Hence, a more sophisticated model would be required to model the behaviour of these critical points lying higher in energy. Further, for a band gap renormalization, one would expect a symmetric shift around the Fermi energy, i.e. the conduction band shifting down and the valence band shifting up by the same amount. In the presented model this means one would need to introduce more fitting parameters to account for the different sign of shifts for different energies. Also the contributions of spin-orbit splitting would need to be taken care of, i.e. complex shifts in the region where the CB of one spin-orbit state overlaps with the VB of the other. Since the VB state blocking signal emerges in a relatively flat region of the static absorbance between 28 and 29 eV, these shifts of the underlying absorbance with the wrong sign only lead to small errors with less than 5% magnitude (see slight blueish colour code between 28 and 29 eV in main text Fig. 2e). This is also the reason as to why in the raw measured transient absorption trace (main text Fig. 2a) the VB state blocking signal can be appreciated as a separate feature.

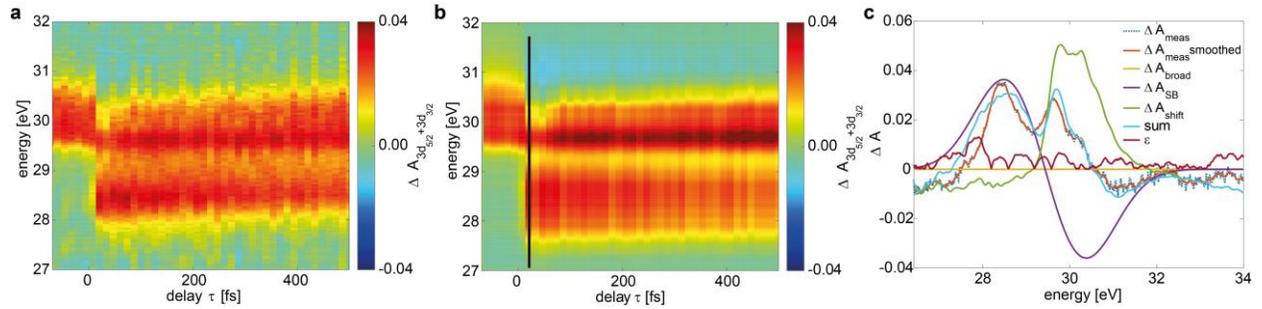

*Figure S3: Comparison of measured differential absorption trace with first iteration of the decomposition model. During the first iteration two Gaussian distributions with magnitudes comparing to the 28.3 eV measured signal are assumed and $\Delta E_{shift}(\tau)$ and $\sigma(\tau)$ are retrieved for each time step. Comparing the measured differential absorption trace (a) with a forward calculated trace (b), i.e. $\Delta A_{SB,0}(E, \tau) + \Delta A_{broad,n}(E, \sigma(\tau)) + \Delta A_{shift,n}(E, \Delta E_{shift}(\tau))$, already shows reasonable agreement. Investigating the components for a single time step in panel (c) shortly after time zero (black line in b) shows that the first guess of Gaussian distributions for the state blocking ($\Delta A_{SB}$) and an absorption change induced by the red-shift of the spectrum by 110 meV ($\Delta A_{shift}$), which comprises heat-induced and photon-induced band shifts and possibly carrier-induced core level shifts, are the main contributions that minimize the error ($\epsilon$) between the measured ($\Delta A_{meas}$) and forward calculated signal (sum).*

## S4 Separation of spin-orbit components

The measured transient absorption of Ge at the $M_{4,5}$-edge as a function of energy, $\Delta A_{\text{meas}}(E)$, is contributed by the transitions from the $3d_{3/2}$ and $3d_{5/2}$ levels, split by 0.58 eV (ref. 8), to the valence and conduction band. Because the core-level states are localized with fixed energy, one can approximate $\Delta A_{\text{meas}}$ as

$$\Delta A_{\text{meas}}(E) = 6\Delta A(E) + 4\Delta A(E - 0.58), \tag{9}$$

where $\Delta A(E)$ is the contribution of a single spin-orbit split core-level state while 6 and 4 are the degeneracy of the $3d_{3/2}$ and $3d_{5/2}$ levels, respectively. It is assumed that the transition strengths follow those degeneracies. Transforming $\Delta A_{\text{meas}}(E)$ into its Fourier counterpart $\Delta \tilde{A}_{\text{meas}}(\eta) = \int e^{-i\eta E} \Delta A_{\text{meas}}(E) dE$, the previous equation becomes

$$\Delta \tilde{A}_{\text{meas}}(\eta) = \Delta \tilde{A}(\eta) \times \left(6 + 4e^{-0.58i\eta}\right). \tag{10}$$

The shifted and factorized contribution of the one spin-orbit split state relative to the other manifests itself as a constant and known phase factor, which can be divided out. By inverse Fourier transform, the contribution of the signal from a single spin-orbit core-level state is retrieved as

$$\Delta A(E) = \int \frac{\Delta \tilde{A}_{\text{meas}}(\eta) \, e^{i\eta E}}{6 + 4e^{-0.58i\eta}} d\eta. \tag{11}$$

The scheme of spin-orbit decoupling is illustrated in Fig. S4. Figure S4a shows the measured data $\Delta A_{\text{meas}}(E)$ versus pump-probe delay, where the features from the two spin-orbit contributions overlap together. Fig. S4b illustrates how the same but factorized signal from two shifted states add up causing a broader feature. Figure S4c is the spin-orbit separated data containing only the excitation from the $3d_{5/2}$ level. The features clear up and more details become visible.

The separation method described here is sensitive to the degeneracy of the spin-orbit states and the spin-orbit energy splitting. Due to the involvement of an imaginary component in the Fourier transform, artefacts will arise as non-zero imaginary parts in $\Delta A(E)$. Hence, the zero value of the imaginary part in $\Delta A(E)$ signifies the success in decoupling the contribution of a single spin-orbit state with this method. In the presented experiment it was useful to retrieve the state blocking contribution to the measured differential absorption signal first (cf. Section S3) and apply the spin-orbit separation only to the state blocking contribution to extract time dynamics of the holes and electrons.

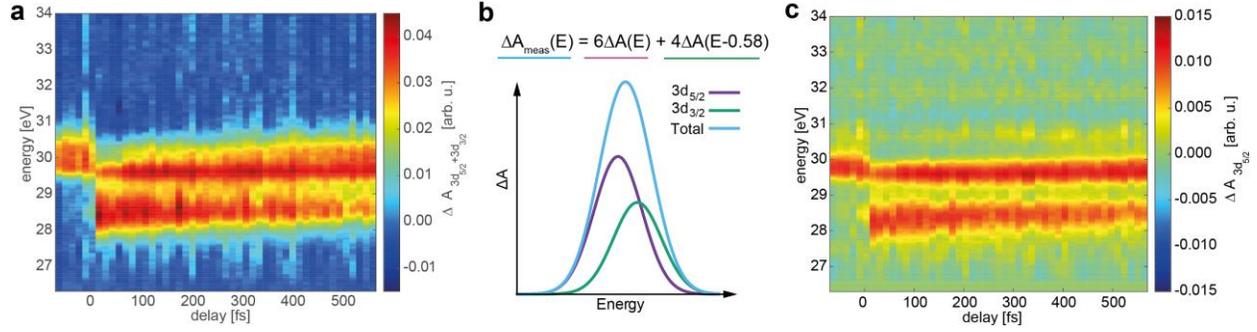

*Figure S4:* Demonstration of decoupling of spin-orbit components. (a) The $3d_{5/2}$ and $3d_{3/2}$ states in germanium cause smeared out transient absorption traces since the spin-orbit splitting of $\Delta E_{so}=0.58$ eV nearly matches the band gap. (b) The measured signal $\Delta A_{meas}(E)$ is the result of the addition of copies of the same signal $\Delta A$ shifted by the spin-orbit splitting and factorized by the degeneracy of the states (6 and 4 for the $3d_{5/2}$ and $3d_{3/2}$ states in germanium, respectively). (c) Using the Fourier method described in the text, one can retrieve a transient absorption trace separated for a single spin-orbit split state.

## S5 Identification of a heat-induced feature in transient absorption data

At a repetition rate of 100 Hz, the germanium sample has 10 ms to relax between consecutive laser pulses, which is considered sufficient for electronic relaxation. Still a transient feature is observed for negative time delays (see Fig. S5a), which can be explained by heat accumulated in the sample during exposure. The 100 nm thick germanium film on a 30 nm thick silicon nitride membrane is thermodynamically nearly a two-dimensional object, and heat can only dissipate in the plane to the thicker silicon wafer that frames the membrane. Measuring the negative time delay feature, which we refer to as heat-induced feature, depending on the pump energy, or average power (Fig. S5a), reveals a near-linear increase of the signal with average power. The origin of the heat-induced feature can be understood by noting that differential absorption is measured, which inherently is sensitive not only to absolute signal changes but also to energy shifts of the absorbance. Fig. S5b shows a differential absorption signal (solid line) calculated from the static absorbance measured on the germanium thin films (see main text Fig. 1d) and a copy of the same signal shifted by 72.8 meV in comparison to the measured heat-induced feature (dotted line) at 0.39 mW average power in the pump arm. The agreement of amplitude and shape confirms that the feature to first order originates from a heat-induced band shift induced by the increased temperature. Using an empirical formula for temperature induced redshift of the conduction band[9]

$$\Delta E_{\text{indirect gap}} = a - \frac{\alpha T^2}{\beta + T} \tag{12}$$

with the parameters $a = 0.741 \text{eV s}$, $\alpha = 4.561 \times 10^{-4} \text{eV/K}$ and $\beta = 210 \text{K}$ for the lowest lying critical point in the conduction band of germanium allows an estimate the temperature of the thin film for different average pump powers by extracting the induced energy shift from the measured heat-induced feature (Fig. S5c). Note that we chose to neglect core-level shifts here, which is reasonable because in contrast to carrier-induced band shifts, where small core-level shifts are expected (cf. Sec. S3), the heat-induced band shifts are expected to have only little impact on the core-level screening. It can be seen that the residual temperature of the thin film is significantly elevated over room temperature even at fractions of a milliwatt average power. From experimental observation, laser annealing sets in at temperatures over 500 K. At pump energies larger than 6 µJ, i.e. 0.6 mW average power, permanent damage of the thin films was observed either by complete destruction of the thin film or by changed static absorbance, indicating a change of structure. To confirm the observation further, a simulation of the heat dissipation using COMSOL was performed. A continuous-wave source of the same average power and 100 µm diameter (FWHM) was assumed for a model of the germanium-silicon-nitride layer system. The three dimensional heat diffusion and radiation losses were considered. Plotted in the inset in Fig. S5c are the retrieved peak temperatures at the excitation spot in comparison to the measured temperatures. The heat diffusion is modelled by a finite element method (COMSOL) with heat conductivity of Ge taken as 5.7 W/(m × K) (ref. 10). The slight deviation in slope can be explained by differing heat conductivities from the literature value[10] for nano to polycrystalline germanium and the grain size in the particular samples. The COMSOL simulation further revealed that the temperature in the hot spot builds up within the first few laser pulses. Hence, it is a good assumption that the data presented in this paper were collected at a given temperature and the temperature gradient during the measurement was marginal.

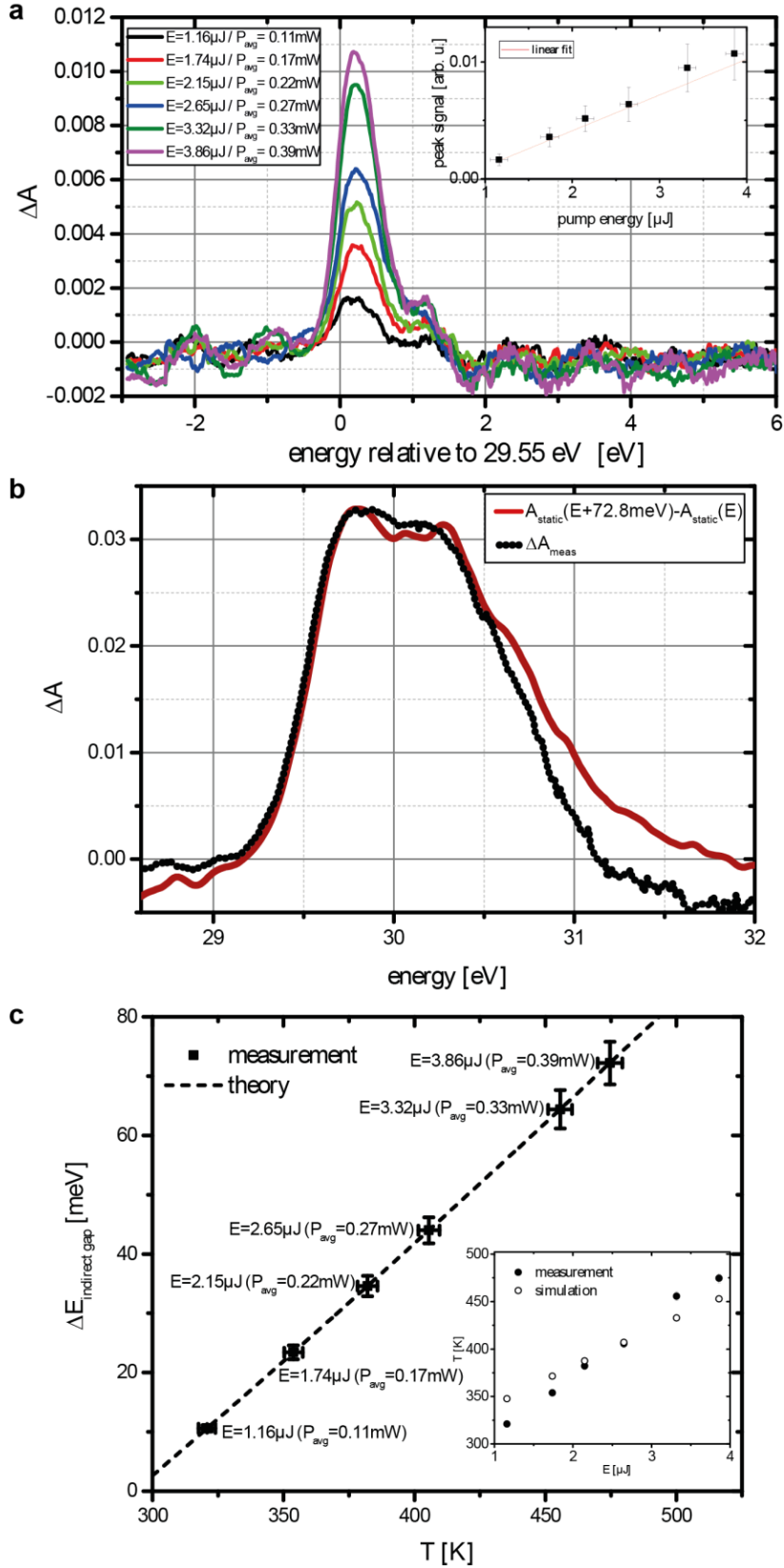

*Figure S5*: Negative time delay feature manifesting as temperature induced shift. (a) For increasing pump energy or average power the heat feature increases nearly linearly. (b) The origin can be explained by a heat-induced band shift, which is confirmed by comparing the measured heat-induced feature with a differential absorption calculated from the measured absorption of the thin film and a shifted copy. In this example for 0.39 mW average power best agreement is achieved for a shift of 72.8 meV. (c) From the measured shift the temperature of the thin film can be derived by using an empirical formula for the redshift. The result is compared to a heat dissipation simulation using COMSOL (inset in c).

The deviations on the high energy side in Fig. S5b can be explained by different temperature coefficients for the higher energy critical points in the conduction band[9]. Hence, for full replication of the heat feature not only a linear shift but also higher order shifts causing modification of the absorbance at elevated temperature must be considered. The linear shift used here can be understood as a first order term for describing the band shift, which shall be sufficient for retrieving the temperature of the thin film and explaining the origin of this feature.

It is important to note that in the experiments always pump off, i.e. cold membrane, was measured first, followed by the pump on measurement, i.e. hot membrane. At each time delay point 100 laser pulses are accumulated for the pump off and pump on conditions, followed by moving a gas cell into the laser beam and the germanium sample out. The gas reference measurements were performed with the same number of laser pulses. Raster scanning was performed on a regular grid of illumination points spaced by 150 microns in both directions. These illumination points were randomly addressed, while the time delay was homogenously increased from negative to positive time delays with the step sizes indicated in the Methods section of the main text. These steps are performed to assure that the temperature of the thin film is relaxed at each new time delay.

## S6 Calculation of the induced band gap renormalization

For a given concentration of free carriers $N_e$ the carrier-induced band gap renormalization, i.e. the combined amount of blueshift of the VB and redshift of the CB, can be calculated by[11]

$$\Delta E_{\text{gap}} = -\left(\frac{e}{2\pi\varepsilon_0\varepsilon_s}\right)\left(\frac{3}{\pi}\right)^{\frac{1}{3}} N_e^{\frac{1}{3}}, \tag{13}$$

where $\varepsilon_0$ is the permittivity of free space and $\varepsilon_s$ is the relative static dielectric constant of the semiconductor. In this work $\varepsilon_{s,Ge} = 15.8$ is used[12]. For $N_e$ values, the were derived by the TDDFT calculation including *k* space response to the used VIS-NIR pulses. The carrier-induced band gap renormalization scales with the third root of the number density of free carriers, due to the average interparticle spacing in three dimensions. In this work only the redshift of the CB is predominantly observed and retrieved by the model outlined in Sec. S3. Thus in the main text the time-dependent redshift of the CB is compared to $\Delta E_{\text{gap}}/2$, which we refer to as band shift rather than a renormalization.

## S7 Sample preparation and characterization

The germanium sample is prepared by electron beam physical vapour deposition (Lebow Company). One hundred nanometre germanium thin films are deposited onto 30 nm thick silicon nitride windows. The germanium film is subsequently annealed at 450 °C for 3 hours. X-ray diffraction (XRD) patterns were acquired using a Bruker GADDS Hi-Start D8 diffractometer with a Co anode at 45 kV/35 mA ($\lambda$=1.79 A). The peak positions and peak widths of the pattern were determined using a commercial software package (Philips X'Pert HighScore Plus). The XRD pattern of Ge was simulated using the peaks (111, 022, 113, and 222) and those were well matched with reference pattern of cubic Ge (ICSD# 98-060-0744). The average grain size (~11 nm) was estimated from X-ray peak broadening using the Scherrer equation after correction for instrumental broadening.

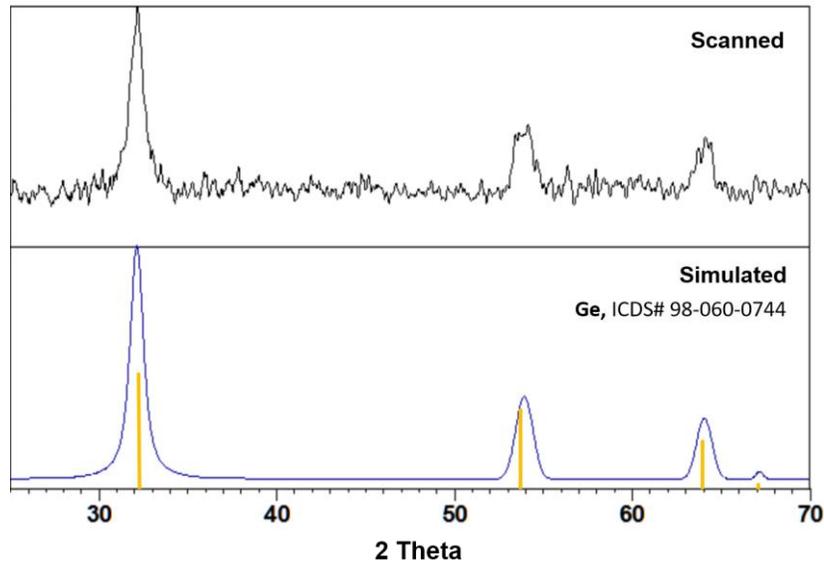

*Figure S6:* XRD pattern collected from the Ge film grown on $Si_3N_4$ window (upper) and simulated pattern from the picks fitted to the XRD pattern (lower).

## S8 Intensity calibration for the pump pulse

To obtain the peak intensity of the pump pulse, we measured the pulse energy $E_p$ at the focus. The pump beam profile features a central spot and a ring originating from diffraction at the hole of the annular mirror that is used for collinear recombination with the XUV. Assuming a Gaussian beam profile at the central spot, the peak intensity of the excitation pulse in vacuum can be written as

$$I_{\text{vac}} = \frac{4\sqrt{\ln 2}}{\pi\sqrt{\pi}} \frac{f_b f_t E_p}{w_x w_y \tau_p}, \tag{14}$$

where $w$ are the half beam diameter for the long and short axes of the elliptical central spot, $\tau_p$ is the excitation pulse duration, $f_b$ is the fraction of pulse energy in the central spot, and $f_t$ is the fraction of energy of the main pulse in the temporal domain.

The temporal profile of the pump pulse is characterized by dispersion scan (D-Scan, Sphere Photonics)[1]. The pulse characterization reveals the pump pulse duration at ~7 fs and energy fraction in the main pulse $f_t \approx 73\%$.

The optical intensity inside the sample $I_s$ is calculated from the vacuum intensity by taking into account the surface reflection by Fresnel equations and the dielectric function of the material. With the index of refraction of germanium at the central wavelength of the pulse, $n(790 \text{ nm}) = 4.7$, the intensity inside the sample can be written as[13]

$$I_s = n\left(\frac{2}{1+n}\right)^2 I_{\text{vac}} \approx 0.6 I_{\text{vac}}. \tag{15}$$

## S9 Orbital character of the CB and VB states

The orbital character for the CB and VB is depicted in the main text, Fig. 1b. Extensive hybridization of the atomic orbitals in the solid band structure leads to most of the conduction band having some percentage of *4p* character, even in regions such as the Γ and L CB valley that are primarily of *4s* character. In the valence band, the heavy- and light- hole bands are of similar energy, except where they bifurcate in the X valley. The split-off band is lower in energy and has more *4s* character. Previous reflection spectroscopy measurements have identified the structure of the $M_{4,5}$-edge at 30 eV as pertaining to the Δ valley[14], but direct excitation to the Δ valley would require two near infrared photons. While such processes are indeed possible at the intensities used in the experiment, the cross-section for two-photon processes is considerably lower. A calculation of the projected density of states reveals that the Γ and L valleys are as much as 30% *4p* character. These regions are thus visible in the core-level spectroscopy experiment, but it can be expected that near the band edge the sensitivity is decreased. This could explain why in Fig. 3a, near the assignment of the $L_1$ valley, the measured signal, although not zero, is weaker than expected.

## S10 Details on the first principle calculations

Real-time TDDFT as implemented in the real-space grid based Ab-initio Real-time Electron Dynamics simulator (ARTED)[15] code is used to simulate the excited-state carrier distribution created in the valence and conduction bands by interaction with a few fs, 780 nm pulse with a peak intensity of $1\times10^{11}$ W/cm$^2$ and the time profile shown in Fig. S7.

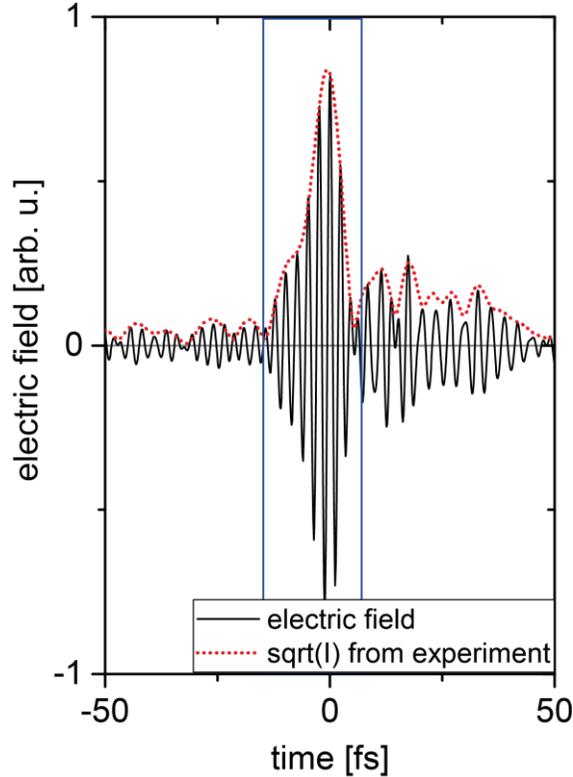

*Figure S7:* *Electric field used for calculations modelling the VIS-NIR excitation pulse used in the experiment. The envelope of the pulse was measured by dispersion scan technique (dotted red line). The electric field is derived using the carrier-frequency of the pulse corresponding to a central wavelength of 780 nm (black solid line). For the calculations a time window of 20 fs width (blue rectangle) around the centre of the pulse was used.*

The temporal pulse profile is derived from experimentally measured intensity and spectral phase information using the dispersion scan technique[1]. The asymmetry of the electric field arises from a residual third order phase on the pulse. From the electric field in Fig. S7 a 20 fs section around the peak electric field (blue rectangle in Fig. S7) was used for the calculations outlined here. Temporal windowing is necessary, since it is more computationally expensive to run longer segments and given the weak intensity outside the central segment it is not expected that significant population differences arise by excluding those. The real-time TDDFT calculations are based on a 64 atom supercell of Germanium and employed norm-conserving pseudopotentials with 4 electrons in the valence. Wavefunctions are expanded on a 32x32x32 real-space grid and a uniform Γ-centred 8x8x8 grid is used for Brillouin-zone integration. The Tran-Blaha-Becke-Johnson potential is employed to describe exchange-correlation effects[16]. The subsequent occupation constrained eXcited-electron Core-Hole (XCH)[38] calculations are based on a 64 atom supercell of Ge, and employ a plane wave basis set with an energy cut-off of 40 Ry and the same 8 x 8 x 8 k-point grid as the TDDFT calculations. Exchange-correlation effects are described at the level of the Perdew-Burke-Ernzerhof[17] functional with a scissors correction employed to recover the full bandgap of germanium.